\documentclass[11pt]{article}
\usepackage{epsfig}
\usepackage{amsfonts}
\usepackage{amsmath}
\usepackage{amssymb}

\newcommand{\ep}{\epsilon}

\newcommand{\al}{\alpha}
\newcommand{\bt}{\beta}
\newcommand{\g}{\gamma}
\newcommand{\ta}{\theta}

\newcommand{\si}{\sigma}

\newcommand{\simu}{\sigma^{\mu\nu}}
\newcommand{\Fmu}{F_{\mu\nu}}

\newcommand{\MT}{M_{\slashTsub}}
\newcommand{\MW}{M_{\mathrm{EW}}}
\newcommand{\slashT}{\slash\hspace{-0.5em}T}
\newcommand{\slashTsub}{\slash\hspace{-0.4em}T}

\newcommand{\slashPT}{\slash\hspace{-0.6em}P\slash\hspace{-0.5em}T}
\newcommand{\slashPTsub}{\slash\hspace{-0.45em}P\slash\hspace{-0.4em}T}

\newcommand{\Nb}{\bar N}
\newcommand{\Fp}{F_\pi}

\newcommand{\tb}{\bar \theta}

\newcommand{\mpi}{m_{\pi}}
\newcommand{\MQCD}{M_{\mathrm{QCD}}}

\newcommand{\Or}{\mathcal O}

\newcommand{\vL}{\ensuremath{\mathcal{L}}}
\newcommand{\vp}{\varphi}
\newcommand{\sq}{^{2}}
\newcommand{\ga}{\gamma}

\newcommand{\dslash}[1]{#1 \llap{/\kern-0.5pt}}
\newcommand{\Dslash}[1]{#1 \llap{/\kern+1.2pt}}
\newcommand{\DDslash}[1]{#1 \llap{/\kern+2.3pt}}
\newcommand{\dslashh}[1]{#1 \llap{/\kern+1pt}}

\newcommand{\boldtau}{\mbox{\boldmath $\tau$}}
\newcommand{\boldpi}{\mbox{\boldmath $\pi$}}

\newcommand{\CP}{C\hspace{-.5mm}P}
\newcommand{\CPT}{C\hspace{-.5mm}PT}                          %

\newcommand{\bea}{\begin{eqnarray}}
\newcommand{\eea}{\end{eqnarray}}
\newcommand{\bma}{\begin{pmatrix}}
\newcommand{\ema}{\end{pmatrix}}
\newcommand{\nn}{\nonumber}

\begin{document}

\begin{titlepage}

\vspace{2.0cm}

\begin{center}
{\Large\bf 
Renormalization Group Running of Dimension-Six\\ 
Sources of Parity and Time-Reversal Violation}

\vspace{1.7cm}

{\large \bf   W. Dekens$^1$ and J. de Vries$^{1,2,3}$} 

\vspace{0.5cm}

{\large 
$^1$ 
{\it KVI, Theory Group, University of Groningen,\\
9747 AA Groningen, The Netherlands}}

\vspace{0.25cm}
{\large 
$^2$ 
{\it Nikhef, Science Park 105, \\ 
1098 XG Amsterdam, The Netherlands}}

\vspace{0.25cm}
{\large 
$^3$ 
{\it Institute for Advanced Simulation, Institut f\"ur Kernphysik, and J\"ulich Center for Hadron Physics, Forschungszentrum J\"ulich, \\
D-52425 J\"ulich, Germany}}

\end{center}

\vspace{1.5cm}

\begin{abstract}
We perform a systematic study of flavor-diagonal parity- and time-reversal-violating operators of dimension six which could arise from physics beyond the SM. We begin at the unknown high-energy scale where these operators originate. At this scale the operators are constrained by gauge invariance which has important consequences for the form of effective operators at lower energies. In particular for the four-quark operators. We calculate one-loop QCD and, when necessary, electroweak corrections to the operators and evolve them down to the electroweak scale and subsequently to hadronic scales. We find that for most operators QCD corrections are not particularly significant. We derive a set of operators at low energy which is expected to dominate hadronic and nuclear EDMs due to physics beyond the SM and obtain quantitative relations between these operators and the original dimension-six operators at the high-energy scale. We use the limit on the neutron EDM to set bounds on the dimension-six operators. 

\end{abstract}

\vfill
\end{titlepage}
\section{Introduction}
The search for flavor-diagonal $\CP$ violation has put stringent bounds on possible $\CP$-violating physics beyond the Standard Model (SM). Electric dipole moments (EDMs) of particles, nucleons, nuclei, atoms, and molecules, which break parity ($P$) and time-reversal invariance $T$ and, by the $\CPT$ theorem, $\CP$ invariance, are the archetypical observables being looked for. However,  despite impressive measurements, so far without success. EDMs are such good probes of physics beyond the SM because, at the current experimental accuracy, the known source of $\CP$ violation in the SM, the phase in the quark-mixing matrix, provides a negligible background \cite{Pospelov:2005pr}. EDMs are, however, not ``SM-free'' probes of new physics because of the existence of a $P$- and $T$-violating ($\slashPT$) interaction in QCD \cite{'tHooft:1976up}. This interaction, parametrized by an angle $\tb$, is flavor diagonal and generates in principle large hadronic and nuclear EDMs \cite{CDVW79} such that the null-measurement of the neutron EDM forces $\tb \leq 10^{-10}$ \cite{dnbound}. This extreme suppression begs for a satisfying explanation which is currently lacking although some solutions have been proposed, for example the Peccei-Quinn mechanism \cite{Peccei:1977hh}. The smallness of $\tb$ leaves room for  $\slashPT$ sources from physics beyond the SM. These sources are the focus of this article while the $\tb$ term will not be considered.

The search for hadronic and nuclear EDMs has grown into an active field of experiments on many different systems \cite{NaviliatCuncic:2012zz}. Ongoing experiments on the neutron EDM aim to improve the upper bound by one or two orders of magnitude \cite{nEDMexp}. Similar progress is expected for the EDMs of diamagnetic atoms such as ${}^{199}$Hg (which currently puts the strongest limit on the proton EDM \cite{hgbound}) and several isotopes of Rn and Ra. Furthermore, there are plans to measure the EDMs of the proton, deuteron, and helion (the nucleus of ${}^3\mathrm{He}$) directly in dedicated storage rings \cite{storageringexpts}. These experiments have an expected accuracy which exceeds the current neutron EDM limit by two to three orders of magnitude. Experiments on all these different systems are needed in order to identify the fundamental $\slashPT$ source. A single measurement can always be reproduced by the $\tb$ term or any new $\slashPT$ hadronic source.

While EDM experiments are taking place at the low-energy frontier, other tests of the SM are being performed at the highest scales at the LHC. Bounds on parameters appearing in various SM extensions are being set and are constantly improving. In this work we wish to set up a framework in which these complementary tests can be compared. An ideal method to build such a framework is the use of effective field theory (EFT) and renormalization-group equations (RGEs). We assume that new physics appears at a scale considerably higher than the electroweak scale. Just below the former scale, after integrating out the non-SM fields, effects of new physics can be described by effective higher-dimensional operators consisting of SM fields and obeying the SM Lorentz and gauge symmetry.  In this way we parametrize our ignorance of the high-energy theory. In principle the list of effective operators is infinite, but they can be ordered by their dimension. The higher the dimension of the operator, the more suppressed its low-energy effects are.  The first $\slashPT$ operators relevant for hadronic and nuclear EDMs appear at dimension six \cite{ Buchmuller:1985jz,  Weinberg:1989dx, DeRujula:1990db, RamseyMusolf:2006vr, Grzadkowski:2010es}. We will not consider (semi-)leptonic dimension-six operators. 

At high energies the dimension-six operators can be measured or bounded directly, while at lower energies, apart from integrating out heavy SM fields, it is necessary to take into account the evolution of the coupling constants and possible mixing of the dimension-six operators. Here these effects are studied simultaneously. We begin right below the unknown scale of new physics and add the dimension-six operators relevant to hadronic and nuclear EDMs. Because the set of operators is large we focus on operators involving first-generation quarks only, leaving the treatment of operators involving heavier quarks to future work. We do consider $\slashPT$ operators containing heavy gauge and Higgs fields which, at low energies, contribute to $\slashPT$ interactions among light fields. 
The operators are evolved, considering one-loop QCD and, when necessary, electroweak corrections, down to the electroweak scale where the $SU(2)_L$ symmetry is spontaneously broken. At the electroweak scale the heavy SM fields decouple from theory and can be integrated out.  We do so and match to an EFT consisting of light fields only. We evolve the resulting operators to a scale around one GeV where QCD becomes nonperturbative (the QCD scale). The QCD running to lower energies has for many operators been calculated in the literature \cite{ Wilczek:1976ry,   BraatenPRL, Degrassi:2005zd, An:2009zh, Hisano:2012cc}. Here we collect, rederive, and, where necessary, supplement  these results. We do the same for electroweak corrections \cite{DeRujula:1990db}.

Our approach differs from previous studies mainly in the focus on gauge symmetry and the systematic treatment of the various dimension-six operators relevant for hadronic and nuclear EDM experiments. By forcing the SM gauge symmetries onto the effective operators it becomes possible to elegantly derive the form of the low-energy $\slashPT$ operators. We will show, for example, that gauge symmetry ensures that only a subset of all possible $\slashPT$ four-quark operators among light quarks \cite{An:2009zh, Hisano:2012cc} will be important at low energies.  Additionally, because the set of effective operators considered at the high-energy scale is very general,  the framework is applicable to many models of new physics. 
Once the matching to the effective operators has been performed, our results can be used to find the restrictions imposed by the EDM limits on the particular high-energy model under investigation.
It is also possible to compare these low-energy bounds with direct bounds obtained from, for example, LHC experiments.

In order to use the hadronic and nuclear EDM limits it is necessary to calculate these quantities in terms of the dimension-six operators at the QCD scale. These calculations are problematic since they involve nonperturbative QCD but reasonable values are obtained by using QCD sum rules \cite{Pospelov:2005pr, Leb04}, chiral perturbation theory \cite{An:2009zh, Vri11a, Mer11, Vri11b, Vri12, Lagdim6}, or naive dimensional analysis \cite{NDA, texas}. Recently, a strategy has been proposed to disentangle the $\tb$ term and dimension-six sources from measurements of the EDMs of the nucleon and light nuclei \cite{Vri11a, Vri12, Bsaisou:2012rg}. This strategy depends on the particular set of dimension-six operators expected to be important at low energies. This set of operators is derived here (a qualitative derivation was given in Ref.\ \cite{Lagdim6}) and can be used as the starting point for calculations of low-energy hadronic $\slashPT$ observables such as EDMs or scattering observables \cite{Song:2011sw}.

Lattice QCD could provide more reliable calculations than chiral techniques and progress has been made in evaluating the nucleon EDM originating in the $\tb$ term \cite{Shintani:2005xg}. Other efforts have focused on the contributions to the nucleon EDM arising from some of the dimension-six operators \cite{latticedim6} and, hopefully, from the others in future calculations. Since lattice calculations are difficult and expensive we suggest to focus on the set of operators obtained here.  

The outline of the article is as follows. In Sec.\ \ref{operators} we list the dimension-six operators right below the unknown scale of new physics. The low-energy effects of some of these operators will be suppressed by small SM factors such as weak gauge couplings or the ratio of quark masses to the electroweak scale. In Sec.\ \ref{unsup} we focus on operators without such suppression and evolve them to the QCD scale, integrating out heavy SM fields in the process. Once the operators are at the QCD scale, we use the strong limit on the neutron EDM to bound the coupling constants of the dimension-six operators. In Sec.\ \ref{sup} we repeat these steps for the operators which do suffer from additional suppression. The limits on these operators will therefore be weaker. We discuss, summarize, and conclude in Sec.\ \ref{conclusion}.

\section{Dimension-six operators}\label{operators}
In this article we assume that physics beyond the SM appears at a scale $\MT$ which, considering the great success of the SM, lies considerably higher than the electroweak scale $M_{\mathrm{EW}} \sim 100\,\mathrm{GeV}$. This assumption implies that no new particles exist with masses around or below the electroweak scale. Just below $\MT$, the effects of new physics can be described by effective higher-dimensional interactions consisting of SM fields only. The interactions result from integrating out the heavier fields appearing in the fundamental theory above $\MT$. These non-renormalizable effective operators have to obey the SM gauge symmetries, otherwise it it hard to understand why the explicit symmetry-breaking does not appear at low energies \cite{DeRujula:1990db}. Around $\MW$, $SU_L(2) \times U_Y(1)$ is spontaneously broken down to $U(1)$ and one is tempted to start the analysis at lower energies by constructing all operators that obey $SU_c(3)\times U(1)$ invariance. The latter strategy does not exploit all available information, as the breaking of $SU_L(2) \times U_Y(1)$ is dictated by the SM and does not happen in some general way. For example, in Refs. \cite{An:2009zh, Hisano:2012cc} the most general  form of $SU_c(3)\times U(1)$ invariant $\slashPT$ four-quark interactions is studied. As will be shown here, generally, only a subset of these interactions actually needs to be considered while others are suppressed due to their particular gauge-symmetry breaking properties.

The lowest-dimensional operators important for flavor-diagonal hadronic $P$ and $T$ violation are of dimension six. The full list of gauge-invariant dimension-six operators was constructed in Refs. \cite{Buchmuller:1985jz, Grzadkowski:2010es}. We focus on $\slashPT$ interactions involving the first generation of quarks only, leaving the inclusion of heavier quarks (and generation-changing operators) to future work. We first define the fields.
The left-handed light-quark fields form a doublet of $SU_L(2)$
\begin{equation}
q_L = \left(\begin{array}{c}
u_L \\
d_L
\end{array}
\right),
\end{equation}
while the right-handed fields $u_R$ and $d_R$ are singlets. The field $\varphi$ denotes an $SU_L(2)$ doublet of scalar fields 
$\varphi^J$, $J=1,2$. It is convenient to define
$\tilde \varphi^I = \varepsilon^{IJ} \varphi^{J*}$,
where $\varepsilon^{IJ}$ is the antisymmetric tensor in two dimensions
($\varepsilon^{12}=+1$). We will adopt the unitarity gauge, in which
\begin{equation}
\varphi  = \frac{v}{\sqrt{2}}  
\left(\begin{array}{c} 0 \\ 1 + \frac{h}{v} \end{array}\right),
\end{equation}
in terms of the vacuum expectation value (vev) of the Higgs field $v$ and the Higgs boson $h$.
The gauge bosons associated with the gauge groups
$SU_c(3)$, $SU_L(2)$, and $U_Y(1)$ are denoted by, respectively,
$G^a_{\mu}$, $W^i_{\mu}$, and $B_{\mu}$.
The covariant derivative of matter fields is
\begin{eqnarray}\label{Cov.1}
D_{\mu} \equiv \partial_{\mu}  - i g_s\, G^a_{\mu} t^a 
- i \frac{g}{2}\, W^i_{\mu} \tau^i - i g^{\prime} Y B_{\mu},
\end{eqnarray}
where $g_s$, $g$, and $g^{\prime}$ are the $SU_c(3)$, $SU_L(2)$, and $U_Y(1)$ coupling constants; and $t^a$ and $\tau^i$ are $SU(3)$ and $SU(2)$ generators, 
in the representation of the field on which the derivative acts. The hypercharge $Y$ for the quark and Higgs fields is given by $Y(q_L)=1/6$, $Y(u_R)=2/3$, $Y(d_R)=-1/3$ and $Y(\varphi)=1/2$. The field strengths are
\begin{eqnarray}
G^a_{\mu \nu} &\equiv& \partial_{\mu} G^a_{\nu} - \partial_{\mu} G^a_{\mu}
+ g_s f^{a b c} G^b_{\mu} G^c_{\nu}, \\
W^i_{\mu \nu} &\equiv& \partial_{\mu} W^i_{\nu} - \partial_{\nu} W^i_{\mu} +
g \varepsilon^{i j k} W^j_{\mu} W^k_{\nu}, \\
B_{\mu \nu} &\equiv& \partial_{\mu} B_{\nu} - \partial_{\nu} B_{\mu},
\end{eqnarray}
with
$f^{abc}$ and $\varepsilon^{ijk}$ denoting 
the $SU(3)$ and $SU(2)$ structure constants.

We now turn to the dimension-six operators.  All coupling constants introduced below are proportional to $M_{\slashTsub}^{-2}$. We start by considering various interactions between a quark, a scalar boson, and a gauge boson
\bea
\vL_{6,qq\vp X } &=& -\frac{1}{\sqrt{2}}\bar q_L \si^{\mu\nu} \tilde\Gamma ^u t^a\frac{\tilde \vp}{v}u_R G_{\mu\nu}^a 
-\frac{1}{\sqrt{2}}\bar q_L \si^{\mu\nu} \tilde\Gamma ^d t^a\frac{ \vp}{v}d_R G_{\mu\nu}^a\nn \\
& & -\frac{1}{\sqrt{2}}\bar q_L \si^{\mu\nu}(\Gamma_{B}^u B_{\mu\nu}+\Gamma_{W}^u \boldsymbol{\tau} \cdot \boldsymbol{W}_{\mu\nu})
 \frac{\tilde \vp}{v}u_R \nn\\
& &-\frac{1}{\sqrt{2}}\bar q_L \si^{\mu\nu}(\Gamma_{B}^d B_{\mu\nu}+\Gamma_{W}^d \boldsymbol{\tau} \cdot \boldsymbol{W}_{\mu\nu})\frac{\varphi}{v}  d_R + \mathrm{h.c.},
\label{dim6edms}
\eea
where $\tilde\Gamma^{u,d}$ and $\Gamma^{u,d}_{B,W}$
are complex coupling constants and an explicit factor $v$ was taken out for later convenience. These coupling constants scale as $M_{\slashTsub}^{-2}$ and because they flip the chirality of the quark field we assume them to be proportional to the SM Yukawa couplings. If this assumption breaks down in particular high-energy models the coupling constants can simply be rescaled. Combined with the Higgs vev, the Yukawa couplings can be traded for the light-quark mass. It is convenient to write Eq.\ \eqref{dim6edms} in terms of the physical gauge-boson fields. Focusing on the $\slashPT$ terms only we find interactions containing massless gauge bosons 
\bea \label{q(C)EDMdef}
\vL_{6,\mathrm{q(C)EDM} } &=&-\frac{1}{2}\big[ m_u Q_u d_u ( \bar u i\simu \ga_5  u)\,eF_{\mu\nu} + m_d Q_d d_d( \bar d i\simu \ga_5  d)\,eF_{\mu\nu} \nonumber\\
 && \qquad + m_u \tilde d_u (\bar u i\simu \ga_5 t^a u)\,G^a_{\mu\nu} + m_d \tilde d_d (\bar d i \simu \ga_5 t^a d)\,  G^a_{\mu\nu}\big]\left(1 + \frac{h}{v}\right),
 \eea
 in terms of the photon field $A^\mu \equiv c_w  B^\mu + s_w W_3^\mu$ and its field strength $F_{\mu\nu} \equiv \partial_\mu A_\nu -\partial_\nu A_\mu$, where $s_w$ ($c_w$) denotes $\sin \theta_w$ ($\cos \theta_w$) where $\theta_w$ is the weak mixing angle. $Q_{u,d}$ denotes the quark electric charge in units of $e = - g s_w = - g^\prime c_w> 0$ and $m_{u,d}$ the quark mass. The terms without scalar fields in the first and second line of Eq.\ \eqref{q(C)EDMdef} can be interpreted as, respectively, the quark electric and chromo-electric dipole moment (quark EDM and CEDM) with coupling constants
 \begin{equation} \label{dq}
d_u \equiv \frac{c_w\mathrm{Im}\,\Gamma_B^u + s_w\mathrm{Im}\,\Gamma_W^u}{e m_u Q_u},\qquad d_d \equiv \frac{c_w\mathrm{Im}\,\Gamma_B^d - s_w\mathrm{Im}\,\Gamma_W^d}{e m_d Q_d},\qquad \tilde d_q \equiv \frac{\mathrm{Im}\,\tilde\Gamma^q}{ m_q}.
\end{equation}

After electroweak symmetry breaking Eq.\ \eqref{dim6edms} also contributes to interactions involving heavy gauge-boson fields. Interactions involving the neutral boson $Z^\mu \equiv c_w  W_3^\mu - s_w B^\mu$, in particular its field strength $Z_{\mu\nu} \equiv \partial_\mu Z_\nu -\partial_\nu Z_\mu$,  are given by
\bea\label{qZEDMdef}
\vL_{6,\mathrm{qZEDM} }
&=& -\frac{g}{2}\big[z_u m_u (\bar u i\simu \ga_5  u)\, Z_{\mu\nu}+z_d m_d  (\bar d i \simu \ga_5 d)\, Z_{\mu\nu}\big]\left(1+\frac{h}{v}\right),
\eea
with coupling constants
\begin{equation}\label{qz}
z_u \equiv \frac{c_w\mathrm{Im}\,\Gamma_W^u - s_w\mathrm{Im}\,\Gamma_B^u}{g m_u}, \qquad
z_d \equiv -\frac{c_w\mathrm{Im}\,\Gamma_W^d + s_w\mathrm{Im}\,\Gamma_B^d}{g m_d}.
\end{equation}
Finally, there are terms involving the charged gauge bosons  $W^\pm_\mu \equiv(W_{1\mu}\mp i W_{2\mu})/\sqrt{2}$ and their field strengths $W_{\mu\nu}^\pm \equiv (W_{\mu\nu}^1\mp i W_{\mu\nu}^2)/\sqrt{2}$,
which are of the form
\bea\label{qWEDMdef}
\vL_{6,\mathrm{qWEDM} }
&=& -\frac{g}{\sqrt 2}m_u w_u\left[(\bar d_L i\simu u_R)\, W_{\mu\nu}^- -i\frac{g}{\sqrt{2}}(\bar u i\simu \ga_5 u)\, W_\mu^+W_\nu^- + \mathrm{h.c.} \right] \left(1+\frac{h}{v}\right), \nn\\
&&-\frac{g}{\sqrt 2} m_d w_d\left[(\bar u_L i\simu d_R)\, W_{\mu\nu}^+ +i\frac{g}{\sqrt{2}}(\bar d i\simu \ga_5 d)\, W_\mu^+W_\nu^-+ \mathrm{h.c.}\right]\left(1+\frac{h}{v}\right),
\eea
where
\begin{equation}\label{qw}
w_u \equiv  \frac{\mathrm{Im}\,\Gamma_W^u }{g m_u}, \qquad w_d \equiv \frac{\mathrm{Im}\,\Gamma_W^d }{g m_d}.
\end{equation}
The quark EDMs and weak EDMs are related by gauge symmetry as can be seen from Eqs.\ \eqref{dq}, \eqref{qz}, and \eqref{qw}.

The next dimension-six operator is the Weinberg operator \cite{Weinberg:1989dx}  which contains only gluons
\bea
\mathcal L_{GGG}= 
\frac{d_{W}}{6} f^{a b c} \varepsilon^{\mu \nu \alpha \beta} 
G^a_{\alpha \beta} G_{\mu \rho}^{b} G^{c\, \rho}_{\nu},
\label{XXX}
\eea
where $\varepsilon^{\mu \nu \alpha \beta}$ is the totally antisymmetric symbol
in four dimensions ($\ep^{0123}=1$). This operator can be interpreted as the CEDM of the gluon \cite{Braaten:1990zt}. 

Next we consider $\slashPT$ interactions among quarks  \cite{RamseyMusolf:2006vr}
\bea\label{qqqq0}
\vL_{6,qqqq} &=& \Sigma_{1}(\bar q^i_L u_R)\varepsilon_{ij}(\bar q^j_L d_R)+ \Sigma_8(\bar q^i_L t_a u_R)\varepsilon_{ij}(\bar q^j_L t_a d_R)+\text{h.c.}
\eea
The $\slashPT$ terms can be rewritten as
\bea\label{qqqq}
\vL_{6,qqqq} &=& i \frac{\mathrm{Im}\, \Sigma_{1}}{2}  (\bar u u \, \bar d \g_5 d+\bar u \g_5 u \, \bar d d -\bar d u \, \bar u \g_5 d-\bar d \g_5 u \, \bar u d)\nonumber\\
&&+ i \frac{\mathrm{Im}\, \Sigma_{8}}{2} (\bar u t^a u \, \bar d \g_5 t^a d+\bar u \g_5 t^a u \, \bar d t^a d -\bar d t^a u \, \bar u \g_5 t^a  d-\bar d \g_5 t^a u \, \bar u t^a d) + \dots,
\eea
where the dots stand for $PT$-even terms. These are the only $\slashPT$ dimension-six four-quark interactions among the first generation of quarks that are allowed by the SM gauge symmetries. These four-quark operators are not suppressed by light-quark masses. We denote the operators in Eq.\ \eqref{qqqq} by FQPS because of their pseudoscalar-scalar form. Based on their color-structure we name the first and second term in Eq.\ \eqref{qqqq}, respectively, the color-singlet and color-octet FQPS operator even though, of course, both operators are scalars under $SU(3)_c$.

At lower energies additional four-quark interactions will appear when the Higgs and heavy gauge bosons that appear in various dimension-six operators are integrated out. An example of such an operator is \cite{Ng:2011ui}
\bea\label{qqWlr}
\vL_{6, qq\vp\vp}& =& \Xi_1\, \bar u_R\ga^\mu d_R(\tilde \vp^\dagger i D_\mu \vp)+\text{h.c.},
\eea
which after electroweak-symmetry breaking becomes
\bea\label{qqW}
\vL_{6, qq\vp\vp}& =& \frac{v^2 g}{2\sqrt 2}\bigg[\Xi_1\,\bar u_R \g^\mu  d_R\,W_\mu^+ + \text{h.c.}\bigg]\left(1+\frac{h}{v}\right)^2.
\eea
Below the electroweak scale, the $W$ boson is integrated out and the terms proportional to $\mathrm{Im}\,\Xi_1$ will contribute to $\slashPT$ four-quark interactions with different structure than Eq.\ \eqref{qqqq} \cite{Ng:2011ui}. The associated factor $G_F$ is compensated by the factor $v^2$ in Eq.\ \eqref{qqW}. 

At the level of dimension-six operators additional interactions among quarks and Higgs bosons appear
\bea\label{qqH0}
\vL_{6,qq\vp\vp\vp}&=&\sqrt{2}\vp^\dagger \vp\left(\bar q_L Y^{\prime u} \tilde \vp u_R+\bar q_L Y^{\prime d}\vp d_R+ \text{h.c.}\right).
\eea
The terms proportional to $(v+h)$ can be absorbed into the Yukawa couplings appearing in the SM. The $\slashPT$ terms that remain are
\bea \label{qqH}
\vL_{6,qq\vp\vp\vp}&=&h\left(v+h\right)\left(v+\frac{h}{2}\right)\left(
i\, \mathrm{Im}\, Y^{\prime u}\,\bar u \g^5 u + i\, \mathrm{Im}\, Y^{\prime d}\,\bar d \g^5 d \right).
\eea

Finally, there are operators consisting purely of gauge and scalar bosons. Analogous to Eq.\ \eqref{XXX} one can write an interaction among $SU(2)_L$ gauge bosons
\bea
\mathcal L_{WWW}= 
 \frac{d_{w}}{6} \varepsilon^{ijk} \varepsilon^{\mu \nu \alpha \beta} 
W^i_{\alpha \beta} W_{\mu \rho}^{j} W^{k\, \rho}_{\nu},
\label{XXX2}
\eea
which can be written in terms of physical gauge bosons as
\cite{DeRujula:1990db}
\begin{eqnarray}\label{Wweakdipole}
-id_{w}\varepsilon^{\mu\nu\al\bt} W_{\bt}^{+\, \rho}W_{\rho\al}^{-}\left(s_w  F_{\mu\nu}+c_w Z_{\mu\nu}-2i g W^+_\mu W^-_\nu\right).
\end{eqnarray}
Other operators involve the Higgs field 
\bea \label{dim6thetaterms}
\vL_{6,XX\vp\vp }&=&  \varepsilon^{\mu\nu\al\bt}\left(\ta ' g_s^2G_{\mu\nu}^a G_{\al\bt}^a+\ta_W 'g^2W_{\mu\nu}^i W_{\al\bt}^i
+\ta_B 'g^{\prime \, 2}B_{\mu\nu} B_{\al\bt} \right)\vp^\dagger \vp\nonumber\\
&&- \varepsilon^{\mu\nu\al\bt}g g'\ta_{WB}' W_{\mu\nu}^{i}B_{\al\bt}\left(\vp^{\dagger}\tau^i\vp\right).
\eea
Without any Higgs field, \textit{i.e.} using $\vp^\dagger \vp / v^2 = 1/2$, the first three operators in Eq.\ \eqref{dim6thetaterms} can be absorbed into the $SU_c(3)$, $SU_L(2)$, and $U_Y(1)$ topological theta terms appearing in the SM. This is not the case for terms which contain at least one explicit Higgs boson. These are given by
\bea \label{dim6thetahiggs}
\vL_{6,XXh\vp }&=& \varepsilon^{\mu\nu\al\bt}\bigg[ \ta ' g_s^2G_{\mu\nu}^a G_{\al\bt}^a + (c_w^2 g^{\prime\,2} \ta_B ' + s_w^2 g^{2} \ta_W ')F_{\mu\nu} F_{\al\bt} + (s_w^2 g^{\prime\,2} \ta_B ' + c_w^2 g^{2} \ta_W ')Z_{\mu\nu} Z_{\al\bt} \nonumber\\
&&- 2 s_w c_w (g^{\prime\,2} \ta_B ' - g^{2} \ta_W ')F_{\mu\nu} Z_{\al\bt} + 2 g^{2} \ta_W'\, W_{\mu\nu}^+ W_{\al\bt}^-\bigg]h(v+\frac{h}{2})+\dots ,
\eea 
where the dots stand for terms involving three gauge bosons which will not be used further on. The fourth term in Eq.\ \eqref{dim6thetaterms} obtains nontopological terms both with and without an explicit Higgs boson \cite{DeRujula:1990db}
\bea\label{thetawb}
\vL_{6,\theta_{WB}}&=&  \varepsilon^{\mu\nu\al\bt}g g'\ta_{WB}'\bigg[ (s_w F_{\mu\nu}+c_w Z_{\mu\nu})(c_w F_{\al\bt}-s_w Z_{\al\bt})h(v+\frac{h}{2})\nn\\
&&-ig (W_\mu^+ W_\nu^-)(c_w F_{\al\bt} - s_w Z_{\al\bt} )(v+h)\sq\bigg].
\eea
Linear combinations of the operators without Higgs bosons in Eqs.\ \eqref{Wweakdipole} and \eqref{thetawb} can be interpreted as the electric dipole and magnetic quadrupole moment of the $W^\pm$ boson \cite{Hagiwara}.

The goal is now to evolve the operators in Eqs.\ \eqref{q(C)EDMdef}, \eqref{qZEDMdef}, \eqref{qWEDMdef}, \eqref{XXX}, \eqref{qqqq}, \eqref{qqW},  \eqref{qqH}, \eqref{dim6thetahiggs}, and \eqref{thetawb} to the low-energy scales $\MW$ and subsequently to $\MQCD$. In the process we have to integrate out the heavy fields and include the effects of QCD and, in some cases, electroweak renormalization-group running. Although all above operators are proportional to two inverse powers of $M_{\slashTsub}$ not all operators will be equally important at low energies. Due to the need to integrate out the heavy gauge and Higgs bosons, certain operators induce low-energy interactions which are, apart from the $M_{\slashTsub}^{-2}$ suppression, additionally suppressed by small SM quantities such as electroweak gauge couplings or the ratio of light-quark masses to the electroweak scale. It is therefore convenient to divide the operators into two sets. The first set contains the operators that induce low-energy interactions without suppression. After studying this set in Sec.\ \ref{unsup}, we will turn in Sec.\ \ref{sup} to the remaining operators which do suffer from additional suppression. 

\section{Unsuppressed operators}\label{unsup}
We begin at the scale $\MT$ where all SM fields are still present in the effective field theory. At this scale, the operators in the first set are given by the quark EDMs and CEDMs in 
Eq.\ \eqref{q(C)EDMdef}, the gluon CEDM in Eq.\ \eqref{XXX}, the four-quark operators in Eq.\ \eqref{qqqq}, the right-handed weak current in Eq.\ \eqref{qqW}, and finally the  quark-Higgs and gluon-Higgs interactions in Eqs.\ \eqref{qqH} and \eqref{dim6thetahiggs}. 
The first goal is to evolve these operators down to the electroweak scale $\MW \sim M_W$, where $M_W$ is the mass of the $W$ boson. At this scale we integrate out the Higgs and heavy gauge bosons. In principle it would be better to integrate out each field at its particular threshold, but this would complicate matters without gaining much. 
Somewhat above $\MW$ we pass the threshold of the top quark which has important consequences for certain operators. 

\subsection{From $\MT$ to $\MW$}\label{sec31}

To calculate the QCD running of the operators it is convenient to write the Lagrangian as
\bea \label{Lsum} \vL_{6}& =& \sum_i C_i(\mu) O_i(\mu), \eea
with $\mu$ the renormalization scale and $O_i(\mu)$ the set of operators
\bea  \label{set1a}
O_{q} &=& -\frac{i}{2}e m_{q} Q_{q} \bar q \simu \ga_5 q\, F_{\mu\nu}, \nn \\
\tilde O_{q} &=& -\frac{i}{2} g_s m_{q}\, \bar q \simu \ga_5 t_a q\, G^a_{\mu\nu}, \nn \\
O_W &=& \frac{1}{6}g_s f_{abc}\varepsilon^{\mu\nu\al\bt}G^a_{\al\bt}G^b_{\mu\rho}G_{\nu}^{c \, \rho}, \nn\\
O_{PS_1} &=& \frac{ig^2_s}{2} (\bar u u \hspace{1mm} \bar d \g_5 d+\bar u \g_5 u \hspace{1mm} \bar d d -\bar d u \hspace{1mm} \bar u \g_5 d-\bar d \g_5 u \hspace{1mm} \bar u d),\nn\\
O_{PS_8} &=& \frac{ig^2_s}{2} (\bar u t^a u \, \bar d \g_5 t^a d+\bar u \g_5 t^a u \, \bar d t^a d -\bar d t^a u \, \bar u \g_5 t^a  d-\bar d \g_5 t^a u \, \bar u t^a d),\nn\\
O_{W_R} &=& \frac{g v\sq }{2\sqrt{2}}(i\bar u_R \g^\mu d_R W_\mu^+ - i\bar d_R \g^\mu u_R W_\mu^- ),\nn\\
O_{qH} &=&v\sq\, h\,\bar q i \g^5 q,\nn\\
O_{g H} &=&  g_s^2 \varepsilon^{\mu\nu\al\bt} G^a_{\mu\nu} G^a_{\al\bt}\,vh,
\eea
where $q \in \{u , d\}$. By comparison with Sec.\ \ref{operators} the coupling constants at $\MT$ are found to be
\bea\label{MT}
C_{q}(\MT) = d_q (\MT),&
\tilde C_q (\MT) = {\displaystyle\frac{\tilde d_q (\MT)}{g_s(\MT)}},\nn\\
C_W(\MT) = \frac{d_W(\MT)}{g_s(\MT)},&\qquad
{\displaystyle C_{\text{PS}_{1,8}}(\MT) = \frac{\text{Im}\, \Sigma_{1,8}(\MT)}{g_s\sq(\MT)}},\nn\\ 
C_{W_R}(\MT) = \text{Im}\, \Xi_{1}(\MT),& \qquad {\displaystyle C_{qH}(\MT)= \text{Im}\,Y^{\prime q}(\MT)},\nn\\
C_{gH}(\MT) = \theta^\prime(\MT),&&
\eea
which are all proportional to $\MT^{-2}$.
We use the following operator basis 
\bea \label{basis}
\vec C (\mu)= (C_{q}(\mu), \tilde C_{q}(\mu), C_W(\mu),  C_{PS_1}(\mu), C_{PS_8}(\mu), C_{W_R}(\mu), C_{qH}(\mu), C_{gH}(\mu))^T .
\eea
The RGE can be expressed as
\bea \label{RGE}
\frac{d \vec C(\mu)}{d \ln \mu} = \g \vec C(\mu),
\eea
where $\g$ is the anomalous dimension matrix which is expanded in powers of $g_s^2/(4\pi)^2 = \alpha_s/(4\pi)$. Up to first order in $\alpha_s$, $\g$ can be written as 
\bea \label{gamma}
\ga = \frac{\al_s}{4\pi} \bma \gamma_{\mathrm{dipole}} & \gamma_{\mathrm{mix}} & 0&0&\gamma_{\mathrm{mix}\,gH} \\
0&\gamma_{PS} &0&0&0\\
0 &0& \gamma_{W_R}&0&0\\
0&0&0&\gamma_{qH}&0\\
0&0&0&0&\gamma_{gH}\ema.
\eea
Most of the submatrices appearing in Eq.\ \eqref{gamma} have been calculated in the literature. The running and mixing of the quark and gluon dipole operators were calculated in Refs. \cite{Weinberg:1989dx, Wilczek:1976ry,  BraatenPRL, Degrassi:2005zd} 
\bea\label{gammadipole} &\ga_{\mathrm{dipole}} = \bma 8 C_ 2 (N) & -8C_ 2 (N)  &0 \\ 0 & 16 C_ 2 (N) - 4 N& 2N \\0 & 0 & N+2n_f+\bt_0\ema,
\eea
where $C_2(N) = (N\sq-1)/2N$ and $\bt_0 = \frac{1}{3}(11N-2n_f)$ in terms of $N$ and $n_f$ the number of, respectively, colors and flavors. Here we confirm these results. 

The submatrix $\gamma_{PS}$ describes the running and mixing of the $\slashPT$ four-quark operators.  In Refs. \cite{An:2009zh, Hisano:2012cc} these matrices were calculated for a different basis of operators consisting of the most general flavor-diagonal $\slashPT$ four-quark operators containing up, down, and strange quarks. In the limiting case of only two flavors, the set contains ten different operators whereas our set, based on gauge invariance, consists of two interactions only. As we will see, below $\MW$ two additional four-quark operators will be induced by $O_{W_R}$, while the remaining six operators are in general suppressed. The suppressed terms are discussed in Sec.\ \ref{sup}.
In the basis of Eq.\ \eqref{basis} the anomalous dimension matrix of the FQPS operators is
\bea\label{gamma4quark}
\ga_ {PS} &=& -2\bma\frac {(3 N + 4) (N\sq - 1)} {N\sq} -\bt_0&\frac {(N + 1)\sq (N - 2) (N - 1)} {N^3} \\
-4\frac {N + 2} {N} &  2\frac {N\sq + 2} {N\sq} - 2 C_ 2 (N)-\bt_0\ema.
\eea
We have checked that Eq.\ \eqref{gamma4quark} agrees with results obtained in Refs. \cite{An:2009zh, Hisano:2012cc} after a suitable basis transformation. 

\begin{figure}[t]
\centering
\includegraphics[scale=0.7]{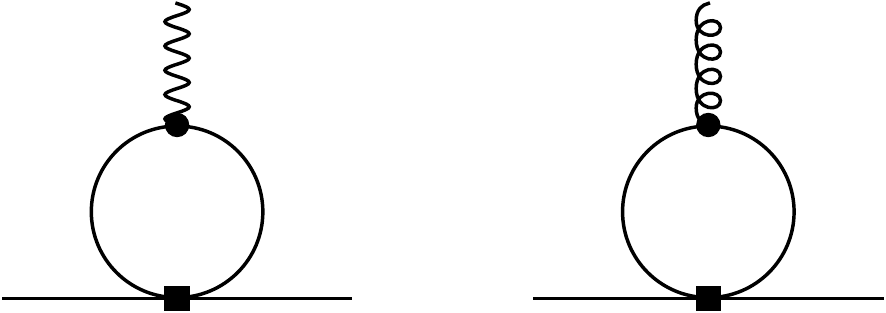}
\caption{One-loop diagrams contributing to the quark electric and chromo-electric dipole moments.
Solid, wavy, and curly lines represent the propagation of
quarks, photons, and gluons respectively.
The square denotes a $\slashPT$ four-quark interaction,
other vertices representing $T$ interactions from the Standard Model.}
\label{FQto(C)EDM}
\end{figure} 

It is necessary to consider mixing of the four-quark interactions with the dipole operators. This is described by the submatrix $\gamma_{\mathrm{mix}}$. The diagrams responsible for this mixing are shown in Fig. \ref{FQto(C)EDM}. The FQPS operators are invariant under chiral symmetry (\textit{i.e.} under global $SU(2)_L\times SU(2)_R$ rotations) \cite{Vri12, Lagdim6} such that they can  only contribute to (C)EDMs, which break chiral symmetry, through a mass insertion. For this reason, the qCEDMs do not induce the four-quark operators and the mixing is one way only. Up to $\Or(\alpha_s)$, the FQPS operators and the gluon CEDM do not mix although they are both chiral invariant. With these considerations we find
\bea\label{gammamix} \ga_{\mathrm{mix}}  &=& 2\frac{m_{q^\prime}}{m_{q}}\bma \frac{Q_{q^\prime}}{Q_{q}} & \frac{Q_{q^\prime}}{Q_{q}}C_2(N)\\ -1 & \frac{N}{2}-C_2(N)\\ 0 &0 \ema ,
\eea
where $q^\prime$ denotes the quark flavor which is different from the flavor of the induced quark (C)EDM. For example, for the mixing between the up quark EDM and $O_{PS_1}$ the relevant entry is $(\ga_{\mathrm{mix}})_{11} = 2(m_d/m_u) (-1/3)/(2/3) =-m_d/m_u$. Again, after a basis transformation, Eq.\ \eqref{gammamix} is in agreement with results in Ref. \cite{Hisano:2012cc}.

The next operator is the right-handed weak current. This operator will obviously not mix with the other operators via QCD corrections due to the presence of the $W$-boson field. At the level of one-loop electroweak diagrams it will induce corrections to the quark (C)EDM, but these corrections are smaller than tree-level diagrams contributing to four-quark operators \cite{Ng:2011ui}. These tree-level diagrams will be considered below. At $\Or(\alpha_s)$ the operator $O_{W_R}$ does not run and its anomalous dimension is zero,
\bea
\g_{W_R}= 0.
\eea

Then there are the $\slashPT$ Higgs-quark interactions with anomalous dimension
\bea\label{gammaqqH}
\g_{q_H} = -6C_2(N).
\eea
At tree- and one-loop level these interactions induce four-quark operators and the q(C)EDM, but such contributions are suppressed by a factor $m_q/v$. At the two-loop level it is possible to circumvent this suppression via a top-quark loop shown in Fig. \ref{BarrZee} which contributes to the qCEDM. This diagram will be discussed below.

\begin{figure}[t]
\centering
\includegraphics[scale=0.7]{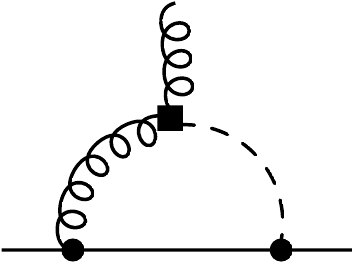}
\caption{One-loop diagram contributing to a quark chromo-electric dipole moment. The dashed line denotes the propagation of a Higgs boson. 
The square denotes the $\slashPT$ gluon-Higgs interaction from Eq.\ \eqref{set1a}. Other notation is as in Fig. \ref{FQto(C)EDM}. For simplicity only one possible ordering is shown here.}
\label{GluonHiggstoCEDM}
\end{figure} 

Finally, we consider the gluon-Higgs interaction. In order to calculate its anomalous dimension and its mixing with the dipoles it is useful to rewrite
\bea
\frac{1}{2}\varepsilon^{\mu\nu\al\bt} G^a_{\mu\nu} G^a_{\al\bt} =\varepsilon^{\mu\nu\al\bt} \partial _\mu \left[G_\nu^a \left(G_{\al\bt}^a-\frac{g_s}{3}f_{abc}G_\al^bG_\bt^c\right)\right],
\eea
which makes it easier to obtain
\bea
\g_{gH} =0,\qquad  \g_{\mathrm{mix}\,gH}= \bma 0& 16& 0  \ema^T,
\eea
where the mixing with the quark CEDM is due to the loop diagram in Fig. \ref{GluonHiggstoCEDM}. The vanishing of  $\g_{gH} $ at the one-loop level is in agreement with Refs.\ \cite{Kaplan:1988ku, Grojean:2013kd}. 

Having calculated the anomalous dimensions of the operators, it is possible to solve the RGE and run the operators down to the electroweak scale $\MW$. During this process we pass the threshold of the top quark. The top quark is integrated out and we match the EFT with six flavors to the EFT with five flavors. As none of the operators contain the top quark the operator basis remains unchanged as we move from one EFT to the next. The matching conditions are 
\bea \vec C^{6}(m_t^+) = \vec C^{5}(m_t^-) ,\eea
where $\vec C^{6}$ are the coupling constants belonging to the EFT incorporating the top quark and $\vec C^{5}$ to the EFT without it and $m_t^+$ ($m_t^-$) denotes the energy scale just above (below) the mass of the top quark. The running is somewhat different for the various EFTs as the anomalous dimension matrix depends on $n_f$ explicitly as well as implicitly through $\al_s(\mu)$. We employ the one-loop running of $\al_s$ for the sake of consistency.
\begin{figure}[t]
\centering
\includegraphics[scale=0.7]{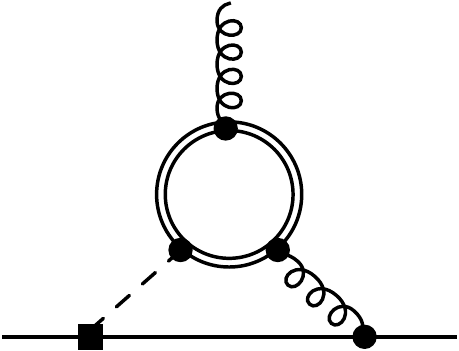}
\caption{Two-loop diagram contributing to the quark chromo-electric dipole moment.
The double solid line denotes a top-quark propagator. 
The square denotes the $\slashPT$ quark-Higgs interaction from Eq.\ \eqref{set1a}. Other notation is as in Figs. \ref{FQto(C)EDM} and \ref{GluonHiggstoCEDM}. For simplicity only one possible ordering is shown here.}
\label{BarrZee}
\end{figure} 
The above treatment works for all operators except for the $\slashPT$ Higgs-quark interactions for which, 
as mentioned above, the passing of the top-quark threshold does have important consequences. The two-loop diagram in Fig. \ref{BarrZee} gives a large contribution to the qCEDM
 \begin{eqnarray}\label{qqHunsup}
\tilde C_q(m^-_t)|_{qq\varphi} = -\frac{ \alpha_s(m_t)}{32\pi^3} \frac{v}{m_q(m_t)}f(\frac{m_t^2}{m_H^2})\,C_{qH}(m_t^+),
\end{eqnarray}
where $f(\frac{m_t^2}{m_H^2})\simeq 1$ \cite{Barr:1990vd} is a function of the top-quark and Higgs mass. 
Here the light quark mass should be evaluated at the scale $m_t$, however, we prefer to present results in terms of quark masses evaluated at the scale $\MQCD$. The one-loop running of the quark masses gives
\bea m_q (m_t)
\simeq 0.55\, m_q(\MQCD).
\label{massRunning}
\eea
Similar diagrams as Fig. \ref{BarrZee} but with gluons replaced by electroweak gauge bosons contribute to the quark EDM, but these contributions are smaller by approximately an order of magnitude due to the smallness of $\alpha_w(m_t)$ compared to $\alpha_s(m_t)$ and we neglect them. Because the two-loop diagram gives by far the most important contribution to low-energy $P$ and $T$ violation coming from Eq.\ \eqref{qqH}, below $m_t$ the quark-Higgs interactions can be neglected. Their effects are saturated by the qCEDM contribution in Eq.\ \eqref{qqHunsup}.
\begin{table}[t]
\caption{\small The size of the quark electric and quark and gluon chromo-electric dipole moments at $M_W$ in terms of various dimension-six operators at $1$ and $10$ TeV. A ``$-$" indicates that there is no contribution at this order.}
\begin{center}\footnotesize
\begin{tabular}{||c|ccccc||}
\hline
    \rule{0pt}{2.5ex}&$ C_q(1 \,\text{TeV})$ & $\tilde{C}_q(1 \,\text{TeV})$ & $C_W(1\, \text{TeV})$ &$ C_{{qH}}(1\, \text{TeV})$&$ C_{{gH}}(1\, \text{TeV})$ \\
 \hline
$ C_q\left(M_W\right)$ & 0.80 &$ 0.18 $& $-0.010$ &
  $-1.4\times 10^{-5} \frac{v}{m_q} $&$-0.032$\\
 $\tilde{C}_q\left(M_W\right) $& $-$ &$ 0.82 $& $-0.090$  & $-2.1\cdot 10^{-4} \frac{v}{m_q} $&$-0.30$\\
$ C_W\left(M_W\right)$ & $-$ & $-$ & $0.64$  & $-$ & $-$ \\  
\hline
  \rule{0pt}{2.5ex} &$ C_q(10\, \text{TeV})$ &$ \tilde{C}_q(10\, \text{TeV})$ &$ C_W(10\, \text{TeV})$  &$ C_{\text{qH}}(10\, \text{TeV})$&$ C_{{gH}}(10\, \text{TeV})$ \\
 \hline
$ C_q\left(M_W\right)$ &$ 0.69$ & $0.26$ & $-0.025 $ &
  $-1.6\times 10^{-5} \frac{v}{m_q}$&$-0.085$ \\
 $\tilde{C}_q\left(M_W\right)$ & $-$ & $0.72$ & $-0.12 $ & $-2.4\cdot 10^{-4} \frac{v}{m_q}$&$-0.48$ \\
 $C_W\left(M_W\right)$ & $-$ & $-$ & $0.46$ & $-$ & $-$\\
 \hline
\end{tabular}
\end{center}
\label{TableMTMW1} 
\end{table}

\begin{table}[t]
\caption{\small The size of the coupling constants of various dimensions-six operators at $M_W$ in terms of the four-quark operators $O_{PS_{1,8}}$ at $1$ and $10$ TeV. A ``$-$" indicates that there is no contribution at this order. The notation $q^\prime$ appearing in the first two rows denotes the quark flavor which is different from the flavor of the induced quark (C)EDM.}
\begin{center}\small
\begin{tabular}{||c|cc||}
\hline
&$ C_{{PS}_1}(1 \,\text{TeV})$ & $C_{{PS}_8}(1\, \text{TeV})$ \\
 \hline
$ C_q\left(M_W\right)$ & $\left(0.0043-0.034 \frac{Q_{q'}}{Q_q}\right) \frac{m_{q^{\prime}}}{m_q} $& $-\left(0.00053+0.043 \frac{Q_{q'}}{Q_q}\right) \frac{m_{q'}}{m_q} $\\
 $\tilde{C}_q\left(M_W\right) $&$ 0.042 \frac{m_{q'}}{m_q}$& $-0.0045 \frac{m_{q'}}{m_q}$\\
$ C_W\left(M_W\right)$ & $-$ & $-$ \\
$ C_{{PS}_1}\left(M_W\right) $& $1.2$ & $0.046$ \\
$ C_{{PS}_8}\left(M_W\right)$ &$-0.26$ & 0.73 \\
 \hline
 &$ C_{\text{PS}_1}(10\, \text{TeV})$ &$ C_{\text{PS}_8}(10\, \text{TeV})$ \\
 \hline
$ C_q\left(M_W\right)$ & $\left(0.012-0.051 \frac{Q_{q'}}{Q_q}\right) \frac{m_{q^{\prime}}}{m_q} $& $-\left(0.0012+0.063 \frac{Q_{q'}}{Q_q}\right) \frac{m_{q'}}{m_q} $\\
 $\tilde{C}_q\left(M_W\right) $& $0.074\frac{m_{q'}}{m_q}$&$-0.0052 \frac{m_{q'}}{m_q}$ \\
$ C_W\left(M_W\right)$ & $-$ & $-$ \\
$ C_{{PS}_1}\left(M_W\right) $& $1.3$ & $0.077$\\
$ C_{{PS}_8}\left(M_W\right)$ & $-0.44$ & $0.58$\\
 \hline 
 
\end{tabular}
\end{center}

\label{TableMTMW2} 
\end{table}

We can now express the dimension-six coupling constants at the electroweak scale ($\MW = M_W$) as function of those at the scale of new physics ($\MT$). Here we consider numerical solutions for two specific values of $\MT$ ($\MT = 1\,\mathrm{TeV}$ and $\MT = 10\,\mathrm{TeV}$). The reason being that the analytical solution of the RGE, while easy to obtain once the anomalous dimensions are known, is too lengthy to write down. The following masses  are used as input \cite{Beringer:1900zz}
\bea
M_Z = 91.2 \, \text{GeV},\qquad M_W = 80.4 \, \text{GeV},\qquad  m_t(m_t) = 160\,\text{GeV},
 \eea
 and for the value of $\al_s$  \cite{Beringer:1900zz}
\bea\label{InputAlpha} \al_s^{n_f=5}(M_Z) = 0.118.\eea
The one-loop running then gives,
\bea\qquad \al_s^{n_f=6}(1\, \text{TeV}) = 0.089, \qquad \al_s^{n_f=6}(10\, \text{TeV}) = 0.073.\eea
We will suppress the $n_f$ dependence of $\al_s$ from here on since the number of flavors is always clear from the context. The results are shown in Tables \ref{TableMTMW1} and \ref{TableMTMW2}. The operator $O_{W_R}$ is not shown because it does not run and gives the trivial result $C_{W_R}(\MT)=C_{W_R}(M_W)$. The results are not very dependent on the choice of $\MT$, because $\al_s$ runs slowly at high energies.

\subsection{From $\MW$ to $\MQCD$}\label{sec32}
The next step is to cross the electroweak scale at which we integrate out the Higgs and heavy gauge bosons. Because the $\slashPT$ Higgs-quark interactions have effectively disappeared from the operator basis around the top-quark threshold, the only operators involving heavy fields are the right-handed weak current and the gluon-Higgs operators. The latter contribute to the quark (C)EDM not only via running from $\MT$ to $\MW$ but also via finite corrections at the Higgs threshold. These corrections, however, appear one order higher in $\alpha_s$ and can be neglected. The  $W$ boson in the right-handed weak current  in Eq.\ \eqref{qqW} can be efficiently integrated out via a tree-level diagram among light quarks \cite{Ng:2011ui}. Since the SM couples the $W$ boson to left-handed quarks only, the appearing effective operator is of the form
\bea
\vL_{6, \mathrm{FQLR}} &=&  i g_s^2 C_{LR_1}\left(\bar u_R \g^\mu d_R\,\bar d_L \g_\mu u_L - \bar d_R \g^\mu  u_R\,\bar u_L \g_\mu d_L \right) + \dots \nonumber\\
&=&
\frac{i g_s^2}{2} C_{LR_1} \left(\bar u \g^\mu \g^5 d\,\bar d \g_\mu u - \bar d \g^\mu \g^5 u\,\bar u \g_\mu d \right) + \dots,
\eea
where the dots stand for higher-dimensional terms appearing from expanding the $W$-boson propagator. The factor $g_s^2$ is introduced for later convenience. The coupling constant right below the electroweak scale is given by
\bea\label{CLR1}
C_{LR_1}(M_W^-) = -\frac{V_{ud}}{g^2_s(M_W)}C_{W_R}(M_W^+) =-\frac{V_{ud}}{g^2_s(M_W)}\,\mathrm{Im}\,\Xi_1(\MT),
\eea
where $V_{ud} \simeq 0.97$ is the up-down CKM element and $M_W^+$ ($M_W^-$) denotes the scale just above (below) the $W$-boson mass. As promised, the factor $v^2$ in Eq.\ \eqref{qqW} has cancelled against the $M_W^2$ appearing in the propagator of the $W$ boson via the SM relation $M_W = g v/2$.  

After integrating out the heavy fields the operator basis has changed. The right-handed weak current, the Higgs-quark, and the Higgs-gluon interactions are removed and two new four-quark operators are added
\bea\label{basisextended}
O_{LR_1} &=&  i \, g^2_s\,(\bar u_{R}\ga^\mu d_{R} \,\bar d_{L}\ga_\mu u_{L}-
\bar d_{R}\ga^\mu u_{R}\, \bar u_{L}\ga_\mu d_{L}),\nn\\
O_{LR_8} &=&  i \, g^2_s\, (\bar u_{R}\ga^\mu t_a d_{R}\, \bar d_{L}\ga_\mu t_a u_{L}-
\bar d_{R}\ga^\mu t_a u_{R}\, \bar u_{L}\ga_\mu t_a d_{L}).
\eea
The coupling constant of $O_{LR_1}$ is given in Eq.\ \eqref{CLR1} while $C_{LR_8}(M_W)=0$. The first operator in Eq.\ \eqref{basisextended} is produced by integrating out the $W$ boson as discussed above, while the second arises when $O_{LR_1}$ is evolved to lower energies. Because these four-quark interactions couple left-handed to right-handed quarks they are denoted by FQLR. The new operator basis becomes
\bea \label{basisMWMQCD}
\vec C (\mu)= (C_{q}(\mu), \tilde C_{q}(\mu), C_W(\mu),  C_{PS_1}(\mu), C_{PS_8}(\mu), C_{LR_1}(\mu), C_{LR_8}(\mu))^T .
\eea
Apart from the appearance of the FQLR operators, the basis has not changed by moving through the $M_W$ threshold, thus for the remaining couplings we have, $C_{i}(M_W^-) = C_{i}(M_W^+)$, $i \neq LR_{1,8}$. The new RGE is written as as
\bea \label{RGE2}
\frac{d \vec C(\mu)}{d \ln \mu} = \g\, \vec C(\mu),
\eea
with the new anomalous dimension matrix
\bea \label{gammab}
\ga = \frac{\al_s}{4\pi} \bma \gamma_{\mathrm{dipole}} & \gamma_{\mathrm{mix}} & 0\\
0&\gamma_{PS} &0\\
0 &0& \gamma_{LR}\ema,
\eea
where, apart from $\g_{LR}$, all entries can be found in the previous section. The diagrams in Fig. \ref{FQto(C)EDM} vanish for the FQLR operators such that they do not induce the quark (C)EDM. Furthermore, the FQPS and FQLR operators do not mix because of their different chiral properties \cite{Lagdim6}, justifying the form of Eq.\ \eqref{gammab}.
The submatrix $\gamma_{LR}$ describes the running and mixing of $O_{LR_1}$ and $O_{LR_8}$. It is given by
\bea\label{gammaFQLR}
\ga_ {LR}= -2\bma -\bt_0  & \frac{3 C_ 2 (N)}{N} \\ 6 & 3\frac {N\sq - 2} {N}-\bt_0\ema,
\eea
which is again in agreement with  Refs. \cite{An:2009zh, Hisano:2012cc} after a suitable basis transformation. 

We can now solve Eq.\ \eqref{RGE2} and run the operators down to the energy scale $\MQCD$. During this process we pass the thresholds of the bottom and charm quarks which are handled the same way as the top-quark threshold. None of the $\slashPT$ operators involve these quarks explicitly such that the operator basis does not change between thresholds. We use \cite{Beringer:1900zz}
\bea
m_b(m_b) = 4.18 \, \text{GeV},\qquad  m_c(m_c) = 1.28 \, \text{GeV}, 
\eea
and run the operators down to $\MQCD=1\,\mathrm{GeV}$. At this scale the strong coupling constant is 
\bea\qquad \al_s(\MQCD) = 0.358.\eea
The results are collected in Tables \ref{TableMWMQCD} and \ref{TableFQLRMWMQCD}. 
\begin{table}
\caption{\small The size of the coupling constants of various dimension-six operators at $\MQCD$ in terms of those of at $M_W$. A ``$-$" indicates that there is no contribution at this order.}
\begin{center}\small
\begin{tabular}{||c|ccccc||}
\hline 
\rule{0pt}{2.5ex} & $C_q\left(M_W\right)$ & $\tilde{C}_q\left(M_W\right) $& $C_W\left(M_W\right) $&$ C_{{PS}_1}\left(M_W\right)$ &$ C_{{PS}_8}\left(M_W\right)$ \\
 \hline
 $C_q\left(M_{\text{QCD}}\right)$ &$ 0.48$ & $0.37 $&$ -0.060 $& $\left(0.040-0.071 \frac{Q_{q'}}{Q_q}\right) \frac{m_{q'}}{m_q}$ &$ -\left(0.0019+0.078\frac{Q_{q'}}{Q_q}\right) \frac{m_{q'}}{m_q} $\\
$ \tilde{C}_q\left(M_{\text{QCD}}\right) $& $-$ & $0.53$ &$ -0.15 $&$ 0.14 \frac{m_{q'}}{m_q}$ &$ -0.0017 \frac{m_{q'}}{m_q}$ \\
$ C_W\left(M_{\text{QCD}}\right) $& $-$ & $-$ & 0.26 & $-$ & $-$ \\
 $C_{{PS}_1}\left(M_{\text{QCD}}\right)$ & $-$ & $-$ & $-$ & $1.5 $& 0.13  \\
$ C_{{PS}_8}\left(M_{\text{QCD}}\right) $& $-$ & $-$ & $-$ & $-0.71 $& 0.28 \\
\hline
\end{tabular}
\end{center}
\label{TableMWMQCD} 
\end{table}
\begin{table}[t]
\caption{\small The size of the coupling constants of the FQLR operators at $\MQCD$ in terms of those at $M_W$.}
\begin{center}
\begin{tabular}{||c|cc||}
\hline 
  &  $C_{LR_1}(M_W)$&  $C_{LR_8}(M_W)$ \tabularnewline
\hline 
$C_{LR_1}(\MQCD)$ & $0.37$&$0.10$\tabularnewline
$C_{LR_8}(\MQCD)$ & $0.47$ &$0.92$  \tabularnewline
\hline
\end{tabular}
\end{center}
\label{TableFQLRMWMQCD} 
\end{table}
By combining Tables \ref{TableMTMW1}-\ref{TableFQLRMWMQCD} it is possible to read off the size the coupling constants at $\MQCD$ in terms of those at $\MT$. Although the bases in Eqs.\ \eqref{basis} and \eqref{basisMWMQCD} are useful to calculate the anomalous dimensions, the explicit appearance of factors of $g_s$ in the operators is unconventional. Here we give the results for the operators without these factors. Defining the quark (C)EDM as in Eq.\ \eqref{q(C)EDMdef} and using  the boundary conditions from Eq.\ \eqref{MT}, we obtain for $\MT = 1 \, \text{TeV}$
\bea \label{q(C)EDMlow1}
 d_q\left(M_{\text{QCD}}\right)&=& 0.39 \,d_q(1 \,\text{TeV})+0.37 \,\tilde{d}_q(1 \,\text{TeV})-0.13\, \theta' \,(1 \,\text{TeV})\nn\\
 && -0.072 \,d_W(1\,
   \text{TeV})-(8.4\cdot 10^{-5})
  \frac{v}{m_q} \mathrm{Im}\,Y^{\prime q}(1 \,\text{TeV})\nn\\
   &&+\left\{0.20,\,0.097 \right\} \mathrm{Im}\,\Sigma _1(1\,
   \text{TeV})+\left\{0.073,\,0.069 \right\}\mathrm{Im}\,\Sigma _8(1\, \text{TeV}),\nn \\
 \tilde{d}_q\left(M_{\text{QCD}}\right)&=& 0.88\, \tilde{d}_q(1\, \text{TeV})-0.34\, \theta' \,(1 \,\text{TeV})\nn\\
 &&-0.29\, d_W(1\, \text{TeV})-(2.4\cdot 10^{-4}) \frac{v}{m_q}
  \mathrm{Im}\,Y^{\prime q}(1 \,\text{TeV})\nn\\
 &&+\left\{0.74,\,0.17 \right\}\mathrm{Im}\,\Sigma
   _1(1 \,\text{TeV})+\left\{0.011,\,0.0025 \right\}\mathrm{Im}\,\Sigma _8(1\, \text{TeV}) ,
   \eea
where the first (second) term in brackets in front of $\mathrm{Im}\,\Sigma _{1.8}$ corresponds to the up (down) quark. For $\MT = 10 \, \text{TeV}$ we find slightly different values
\bea \label{q(C)EDMlow10}
 d_q\left(M_{\text{QCD}}\right)&=&0.33\, d_q(10\, \text{TeV}) +0.41\,\tilde{d}_q(10 \,\text{TeV})-0.22\,\theta' (10 \,\text{TeV})\nn\\
   &&-0.089\, d_W(10\,
   \text{TeV})-(9.5\cdot 10^{-5}) \frac{v}{m_q} \mathrm{Im}\,Y^{\prime q}(10\,
   \text{TeV})\nn\\
   &&+\left\{0.30,\,0.14 \right\} \mathrm{Im}\,\Sigma _1(10\, \text{TeV})+\left\{0.092,\,0.085 \right\}\mathrm{Im}\,\Sigma _8(10 \,\text{TeV}) ,\nn\\
 \tilde{d}_q\left(M_{\text{QCD}}\right)&=& 0.85\, \tilde{d}_q(10 \,\text{TeV})-0.54 \, \theta' (10 \,\text{TeV})\nn\\
 &&-0.30\, d_W(10\, \text{TeV})-(2.7\cdot 10^{-4}) \frac{v}{m_q}
   \mathrm{Im}\,Y^{\prime q}(10 \,\text{TeV})\nn\\
 &&+\left\{1.1,\,0.25 \right\}\mathrm{Im}\,\Sigma
   _1(10\, \text{TeV})+\left\{0.034,\,0.0078 \right\}\mathrm{Im}\,\Sigma _8(10 \,\text{TeV}) .
   \eea
Without looking at specific models of new physics it is difficult to compare the various contributions to the low-energy quark (C)EDMs. Assuming the various coupling constants to be of equal size at $\MT$ and using $v/m_q \sim \Or(10^{5})$ one would conclude that the low-energy quark (C)EDMs are dominated by the $\slashPT$ Higgs-quark interactions. However, in certain models of new physics the Yukawa-like couplings $Y^{\prime u,d}$ scale with the quark mass and are thus $\Or(m_q /vM_{\slashTsub}^2)$ instead of $\Or(1 /M_{\slashTsub}^2)$ as assumed in Sec.\ \ref{operators} \cite{Zhang:2007da}. The resulting suppression makes the contributions proportional to $Y^{\prime u,d}$ negligible. In such a scenario, the low-energy quark EDM gets contributions of roughly equal size from all other dimension-six operators at $\MT$ although the dependence on the gluon CEDM and color-octet FQPS is somewhat smaller.
The low-energy quark CEDMs are mainly determined by the high-energy CEDMs and, depending on the quark flavor, the color-singlet FQPS. The contributions from the gluon CEDM and the gluon-Higgs interaction are also significant, while the influence of the color-octet FQPS can be neglected. We stress again that these statements are very model dependent since the coupling constants at the high-energy scale might, depending on the model of physics beyond the SM, possess a large hierarchy. Some may even be zero. In any case, it is unlikely that all coupling constants at $\MT$ are of similar size.

The next operator is the gluon CEDM, which is defined without $g_s$ as in Eq.\ \eqref{XXX}. At this order, the gluon CEDM does not depend on other sources such that
\bea \label{gCEDMlow}
 d_W\left(M_{\text{QCD}}\right)=0.33 \,d_W\left(1 \, \text{TeV}\right), \qquad d_W\left(M_{\text{QCD}}\right)=0.27 \,d_W\left(10 \, \text{TeV}\right).
\eea
The gluon CEDM is somewhat suppressed at low energies in agreement with Refs. \cite{DeRujula:1990db, BraatenPRL, Degrassi:2005zd}. Additional contributions to the gluon CEDM do appear if  the CEDMs of heavy quarks are taken into account \cite{BraatenPRL}.

The FQPS operators do not mix with other four-quark operators such that the solution of the RGE is very simple. For $\MT = 1 \, \text{TeV}$
  \bea \label{FQPSlow1}
 \mathrm{Im}\,\Sigma_1\left(M_{\text{QCD}}\right)&=&7.2\, \mathrm{Im}\,\Sigma_1\left(1\,\text{TeV}\right)+0.66\, \mathrm{Im}\,\Sigma_8\left(1\,\text{TeV}\right), \nn\\
 \mathrm{Im}\,\Sigma_8\left(M_{\text{QCD}}\right)&=&-3.7\, \mathrm{Im}\,\Sigma_1\left(1\,\text{TeV}\right)+0.69 \,\mathrm{Im}\,\Sigma_8\left(1\,\text{TeV}\right),
 \eea
and for $\MT = 10 \, \text{TeV}$ 
  \bea \label{FQPSlow10}
 \mathrm{Im}\,\Sigma_1\left(M_{\text{QCD}}\right)&=&10\, \mathrm{Im}\,\Sigma_1\left(10\,\text{TeV}\right)+0.95\, \mathrm{Im}\,\Sigma_8\left(10\,\text{TeV}\right), \nn\\
 \mathrm{Im}\,\Sigma_8\left(M_{\text{QCD}}\right)&=&-5.3\, \mathrm{Im}\,\Sigma_1\left(10\,\text{TeV}\right)+0.53 \,\mathrm{Im}\,\Sigma_8\left(10\,\text{TeV}\right).
 \eea 
Again assuming that the coupling constants are of the same order at the high-energy scale, the FQPS operators at $\MQCD$ are dominated by $\mathrm{Im}\,\Sigma_1(\MT)$. Barring unexpected fine-tuning, $ \mathrm{Im}\,\Sigma_1\left(M_{\text{QCD}}\right)$ and  $\mathrm{Im}\,\Sigma_8\left(M_{\text{QCD}}\right)$ are of approximately the same size and both operators should be  taken into account at low energies. It is interesting that the coefficient of the color-singlet operator grows significantly when evolved to lower energies.

Finally, we look at the FQLR operators. Because the operator $O_{W_R}$, from which the FQLR operators originate, does not run between $\MT$ and $\MW$ the results do not depend on the particular value of $\MT$. (Of course, this is true only as far as the running is concerned. The size of the couplings is proportional to $\MT^{-2}$). The results are given for the operators without explicit factors of $g_s$. The redefined couplings are
\bea 
\mathrm{Im}\,\Xi_{LR_1}(\mu)\equiv g_s(\mu)\sq C_{LR_1}(\mu), \qquad \mathrm{Im}\,\Xi_{LR_8}(\mu)\equiv g_s(\mu)\sq C_{LR_8}(\mu),\eea
for which we find
\bea\label{FQLRlow}
\mathrm{Im}\, \Xi_{LR_1}\left(M_{\text{QCD}}\right)=-1.1 \,V_{ud}\, \mathrm{Im}\,\Xi_1\left(\MT\right), \qquad 
 \mathrm{Im}\,\Xi_{LR_8}\left(M_{\text{QCD}}\right)=-1.4 \,V_{ud}\,\mathrm{Im}\,\Xi_1\left(\MT\right).
 \eea
The conclusion is that both four-quark operators are of approximately the same magnitude at low energies even though $O_{\text{LR}_8}$ does not get a direct contribution at the electroweak scale. Clearly, the coupling constants of the two FQLR operators are not independent. Both depend on the same fundamental parameter $\Xi_1(\MT)$.

The right-handed weak current in Eq.\ \eqref{qqWlr} not only gives rise to the FQLR operator but also to $\slashT$ semi-leptonic interactions by coupling it to the left-handed lepton current.  The operator produced in this way does not contribute to hadronic EDMs but, instead, contributes to the $\slashT$ triple correlation $\sim D\,\vec J\cdot(\vec p_e\times \vec p_\nu)$ in nuclear $\beta$-decay \cite{Ng:2011ui}. Both the FQLR operator and its semi-leptonic cousin arise from the same dimension-six operator and depend on the parameter $\mathrm{Im}\,\Xi_1(\MT)$. In Ref.\  \cite{Ng:2011ui} it is argued that the best limit on $\mathrm{Im}\,\Xi_1(\MT)$ comes from the limit on the neutron EDM. Furthermore, the authors of Ref. \cite{Ng:2011ui} used the neutron EDM limit to set a strong constraint on the coefficient $D$. However, QCD corrections were neglected and one might wonder whether their effect is significant. The semi-leptonic operator does not run under QCD RGE which makes it sufficient to consider the running of the FQLR operator. The running of the color-singlet FQLR gives an enhancement of $10\%$ which does not alter the results of  Ref.\ \cite{Ng:2011ui} in a significant way. However, the fact that at low energies not only the color-singlet but also the color-octet operator is present, with  a larger coefficient, may be more important. It is difficult to say how this affects the bound obtained in Ref.\ \cite{Ng:2011ui} since this depends on nonperturbative physics linking the FQLR operators to the neutron EDM. In the event of a partial cancellation between the neutron EDM contributions from the color-singlet and color-octet FQLR operators, the bound on the $D$ coefficient would be weakened. 

To summarize this section we write the $\slashPT$ Lagrangian at the scale $M_{\text{QCD}}$ as
\bea \label{set1MQCD}
\vL_{6, \mathrm{set}\,1} &=& -\frac{1}{2}\bigg(
  d_u e m_{u} Q_{u}\, \bar u i\simu \ga_5  u\,F_{\mu\nu} + d_d e m_{d} Q_{d} \, \bar d i \simu \ga_5 d\, F_{\mu\nu}\nonumber\\
&&+ \tilde d_u  m_{u} \, \bar u i\simu \ga_5 t^a u\,G^a_{\mu\nu} +\tilde d_d m_{d} \, \bar d i \simu \ga_5 t^a d\, G^a_{\mu\nu}\bigg)\nonumber\\
&& + \frac{d_{W}}{6} f^{a b c} \varepsilon^{\mu \nu \alpha \beta} 
G^a_{\alpha \beta} G_{\mu \rho}^{b} G^{c\, \rho}_{\nu} \nonumber\\
 && +i \frac{\mathrm{Im}\, \Sigma_{1}}{2} (\bar u u \, \bar d \g_5 d+\bar u \g_5 u \, \bar d d -\bar d u \, \bar u \g_5 d-\bar d \g_5 u \, \bar u d)\nonumber\\
&&+ i \frac{\mathrm{Im}\, \Sigma_{8}}{2}  (\bar u t^a u \, \bar d \g_5 t^a d+\bar u \g_5 t^a u \, \bar d t^a d -\bar d t^a u \, \bar u \g_5 t^a  d-\bar d \g_5 t^a u \, \bar u t^a d)\nonumber\\
&& +i\,\mathrm{Im}\,\Xi_{LR_1} \left(\bar u_R \g^\mu d_R\,\bar d_L \g_\mu u_L - \bar d_R \g^\mu  u_R\,\bar u_L \g_\mu d_L \right) \nn\\
&& +i\,\mathrm{Im}\,\Xi_{LR_8} \left(\bar u_R \g^\mu t^a d_R\,\bar d_L \g_\mu t^a u_L - \bar d_R \g^\mu t^a u_R\,\bar u_L \g_\mu t^a d_L \right) , 
\eea
where the coupling constants are all evaluated at $\MQCD$. Their values in terms of the coupling constants at the high-energy scale can be read from Eqs.\ \eqref{q(C)EDMlow1}-\eqref{FQLRlow}. We expect the above Lagrangian to capture the most important contributions from physics beyond the SM to first generation $\slashPT$ observables. For example, the EDMs of the nucleon and light nuclei should be calculated as a function of the various coupling constants appearing in Eq.\ \eqref{set1MQCD}. At lower energies, Eq.\ \eqref{set1MQCD} induces $\slashPT$ interactions among pions, nucleons, and heavier baryons. These interactions have been constructed in Ref. \cite{Lagdim6}  by use of chiral perturbation theory.

An important result of this section is the form of the $\slashPT$ four-quark operators. Although they look complicated, their form is very intuitive if one insists on gauge invariance at the scale where the dimension-six operators are generated. This is also indicated by the simple block-diagonal form (apart from mixing with the quark dipoles) of the anomalous dimension matrix. We conclude that around a scale $\MQCD \sim 1\,\mathrm{GeV}$ there are four $\slashPT$ four-quark operators that need to be taken into account with three independent couplings. We will demonstrate in Sec.\ \ref{supqqH} that the remaining six combinations also appear but are, in general, suppressed.
 
\subsection{Bounds from the neutron EDM}\label{bounds}
In this section we use the stringent limit on the neutron EDM, $d_n \leq 2.9\cdot 10^{-13}\,e\,\mathrm{fm}$ \cite{dnbound} to set bounds on the various dimension-six coupling constants. This requires the calculation of the neutron EDM in terms of the couplings appearing in Eq.\ \eqref{set1MQCD} which, in turn, can be related to the couplings at the high-energy scale $\MT$ using the results obtained in the previous sections. Calculating the neutron EDM in terms of the quark-gluon operators is a problematic task due to the nonperturbative nature of QCD at low energies. Despite this difficulty, several approaches exist to tackle this problem (for a review, see Ref. \cite{Pospelov:2005pr}). Here we use recent results obtained in chiral perturbation theory ($\chi$PT) \cite{Vri11a, Mer11, Vri12, Lagdim6}, which has the advantage that all operators appearing in Eq.\ \eqref{set1MQCD} have been treated within the same framework. Furthermore, it allows for the reliable calculation of light-nuclear EDMs \cite{Vri11b, Vri12, Bsaisou:2012rg} which have become the subject of experimental investigation \cite{storageringexpts}.

In $\chi$PT, the low-energy EFT of QCD,  the effective degrees of freedom are pions and nucleons (and heavier baryons) whose interactions are determined by the symmetries of QCD and how they are (spontaneously and explicitly) broken. Pions are interpreted as the Goldstone bosons of the spontaneously broken chiral symmetry of
QCD, $SU_{L}(2)\times SU_{R}(2)$. For a review of $\chi$PT see, for example, Refs. \cite{Weinberg, reviewchiPT}. The extension of $\chi$PT to include the effects of the dimension-six operators has been performed in Ref. \cite{Lagdim6} (the $\tb$ term has been studied in Ref. \cite{BiraEmanuele}). At leading order in the calculation of the neutron EDM only three hadronic $\slashPT$ interactions play a role \cite{Vri11a, Lagdim6} for all dimension-six sources in Eq.\ \eqref{set1MQCD}. These are given by
\begin{eqnarray}
\mathcal{L}_{\slashPTsub}  = 
-\frac{\bar g_0}{F_{\pi}}
\bar{N}\boldtau\cdot\boldpi N - 2\, \Nb\left(\bar{d}_{0}+\bar{d}_{1}\tau_{3}\right)S^{\mu}N\, v^{\nu}\Fmu,
\label{3Lecs}
\end{eqnarray}
in terms of the nucleon doublet $N = (p\,\,n)^T$, the pion triplet $\boldpi$, and the pion decay constant $\Fp = 186$ MeV. In Eq.\ \eqref{3Lecs} the heavy-baryon framework \cite{Jenkins:1990jv, reviewchiPT} has been applied
where, instead of gamma matrices, it is the nucleon velocity $v^\mu$ and spin $S^\mu$
that appear. The first interaction in Eq.\ \eqref{3Lecs} is a chiral-symmetry-breaking $\slashPT$ $\pi N$ interaction, while the other two are short-range, \textit{i.e.} due to dynamics of shorter range than pions, contributions to the isoscalar and isovector nucleon EDM respectively. Which of the hadronic interactions plays a role in the calculation of the neutron EDM depends on the particular dimension-six operator under investigation. 

The actual leading-order calculation of the neutron EDM is fairly straightforward and gives 
\begin{eqnarray}\label{dn}
d_n = (\bar d_0 - \bar{d}_1)
+\frac{eg_A\bar{g}_0}{(2\pi\Fp)^2}
 \ln\frac{\mpi^2}{m_N^2},
\end{eqnarray}
where $g_A \simeq 1.27$ is the strong pion-nucleon coupling constant and $\mpi \simeq 137\,\mathrm{MeV}$ ($m_N\simeq 938\,\mathrm{MeV}$) the mass of the pion (nucleon) \cite{CDVW79}.  Higher-order corrections have been calculated for all dimension-six sources in Refs. \cite{Mer11, Lagdim6}, but they do not change the results significantly. The main difficulty (and uncertainty) in the calculation stems from the estimation of the low-energy constants (LECs) $\bar g_0$ and $\bar d_0 - \bar d_1$ in terms of the couplings in Eq.\ \eqref{set1MQCD}. Here we follow Ref. \cite{Vri11a, Lagdim6} and use naive dimensional analysis (NDA) \cite{NDA, texas}. Hopefully, these estimates will be replaced by lattice-QCD calculations. All estimates below are in terms of the dimension-six coupling constants at the scale $\MQCD$. 

For the quark EDM operators in Eq.\ \eqref{set1MQCD} the main contribution to the neutron EDM comes from the short-range contribution $\bar d_0 - \bar d_1$. The reason being that the non-electromagnetic pion-nucleon interaction is suppressed by the necessity of integrating out the photon appearing in the quark EDM operator. NDA gives $\bar d_0 - \bar d_1 = \Or(e m_{u} Q_{u} d_u,\,e m_{d} Q_{d} d_d)$  consistent with results obtained in quark models \cite{Dib:2006hk}.  For the quark masses we use the values at $\mu = \MQCD$, $m_u(\MQCD) = 3.1\, \text{MeV}$ and $m_d(\MQCD) = 6.5\, \text{MeV}$ \cite{Beringer:1900zz}. 

The quark CEDM operators induce both $\bar g_0$ and, by including $PT$-even electromagnetic effects, $\bar d_0-\bar d_1$. Both contributions in Eq.\ \eqref{dn} appear formally at the same order (one expects, \textit{a priori}, no cancellation among them) but the chiral logarithm, $\log (\mpi^2/m_N^2) \simeq -4$, somewhat enhances the $\bar g_0$ contribution. NDA gives $\bar g_0 = \Or(m_q \tilde d_q \MQCD^2/4\pi)$ consistent with a calculation using QCD sum rules \cite{Pospelov:2001ys}. 

The gluon CEDM conserves chiral symmetry and, as a consequence, its contribution to $\bar g_0$ (the LEC of a chiral-symmetry-breaking interaction) is suppressed by $m_q/\MQCD$ \cite{Demir:2002gg, Lagdim6}. The main contribution to the nEDM stems from the short-range contribution which is estimated by $\bar d_0-\bar d_1 = \Or (e d_W \MQCD/4\pi)$. This estimate is approximately twice as large as a calculation based on QCD sum rules \cite{Demir:2002gg} which reflects the intrinsic uncertainty in these kind of estimations. 

Next are the FQPS operators with coupling constants $\mathrm{Im}\,\Sigma_{1,8}$. As pointed out in Refs. \cite{Vri12, Lagdim6} these operators conserve, just as the gluon CEDM, chiral symmetry such that for these sources $\bar g_0$ is suppressed as well. The short-range terms are estimated as $\bar d_0-\bar d_1 = \Or (e (\mathrm{Im}\,\Sigma_{1,8}) \MQCD/(4\pi)^2)$.

Finally, the FQLR operators do break chiral symmetry and, similar to the quark CEDM, the neutron EDM is determined by both terms in Eq.\ \eqref{dn}. However, for this dimension-six source $\bar g_0$ is smaller than expected (something which goes beyond NDA) due to the smallness of the proton-neutron mass difference (for details, see Ref. \cite{Lagdim6}), such that the main contribution comes again from the short-range terms. NDA gives $\bar d_0-\bar d_1 = \Or (e (\mathrm{Im}\,\Xi_{1,8}) \MQCD/(4\pi)^2)$. This estimate for the neutron EDM is about an order of magnitude smaller than the result obtained in Ref. \cite{An:2009zh}. A possible explanation for this discrepancy is given in Ref. \cite{Lagdim6}. To be on the safe side we will use the smaller estimate. 

\begin{table}[t]
\caption{\small Bounds on the coupling constants of various dimension-six operators at $\MQCD$ and $M_W$ in units of $(100\, \text{TeV})^{-2}$. }
\begin{center}\footnotesize
\begin{tabular}{||c|cc||}
\hline
& $\mu = \MQCD$&$\mu=M_W$ \\
\hline 
  $d_{u,\,d}\left(\mu\right)$ & $\leq \lbrace 7.1,\,6.8 \rbrace$& $\leq \lbrace 15,\,14 \rbrace $\\
  $\tilde{d}_{u,\,d}\left(\mu\right) $& $\leq \lbrace 17,\,8.0 \rbrace $& $\leq \lbrace 18,\,8.7 \rbrace$\\
  $d_W\left(\mu\right) $&$\leq 0.19$&$\leq 0.42$\\
  $\mathrm{Im}\,\Sigma_1 \left(\mu\right)$ &$\leq 2.3$ & $\leq 0.51$\\
  $ \mathrm{Im}\,\Sigma_8\left(\mu\right)$ &$\leq 2.3$ & $\leq 2.8$\\
  $\mathrm{Im}\,\Xi_{LR_1} \left(\mu\right)$ &$\leq 2.3$& $\leq 1.7$ \\
  $ \mathrm{Im}\,\Xi_{LR_8}\left(\mu\right)$&$\leq 2.3$&$\leq0.85$\\
  \hline
\end{tabular}
\end{center}
\label{TableMWMQCDBounds} 
\end{table}

The bounds on the dimension-six coupling constants in Eq.\ \eqref{set1MQCD} are shown in Table \ref{TableMWMQCDBounds} for the scales $\mu = \MQCD$ and $\mu = M_W$. For the latter bounds, certain operators contribute to the neutron EDM in different ways. For example, a quark CEDM at $M_W$ induces both a quark EDM and CEDM at $\MQCD$ which both contribute to the neutron EDM. In these cases we present the strongest bound and do not take into account possible cancellations between different contributions. The limits in Table \ref{TableMWMQCDBounds} involve NDA factors and contain significant uncertainties. All bounds are of order $(100\,\mathrm{TeV})^{-2}$ apart from the bounds on the quark (C)EDMs which are approximately an order of magnitude weaker because they scale with the small quark mass. The running from the electroweak scale to low energies does not affect the bounds by a large amount. For most operators less than a factor two, the exception being the color-singlet FQPS operator whose bound is strengthened by a factor $5$. The QCD corrections are thus not very significant, especially not in the light of the large uncertainties involved. 

It is also possible to set bounds on the couplings at energies above the electroweak scale. However, this depends on the particular value of $\MT$ in two ways. First, the coupling constants are expected to scale as $\sim \MT^{-2}$. We therefore present bounds on the dimensionless combination $\MT^2 C_i(\MT)$ where $C_i$ denotes the various coupling constants. Second, the QCD corrections depend on the particular value of $\MT$ as well. 
In the first two columns of Table \ref{TableMTMQCDBounds} we present bounds for two values of $\MT$. The last three rows contain couplings which did not appear in Table \ref{TableMWMQCDBounds} because the corresponding operators decoupled from the EFT below the electroweak scale. Although the gluon-Higgs operator only contributes to the qCEDM via a one-loop diagram, the bounds on $\theta^\prime$ and $\tilde d_q$ are equally strong.  

It must be stressed that all bounds obtained here cannot simply be used to set bounds on $\MT$. The reason is that the effective couplings can depend on dimensionless factors appearing in the high-energy theory and factors of $4\pi$ arising from integrating out the heavy fields. In fact, as argued in Ref.\ \cite{Arzt:1994gp}, the couplings $d_{u,d}$ $\tilde d_{u,d}$, $d_W$, and $\theta^\prime$ can only be generated at the loop level and are therefore associated with a factor $1/(4\pi)^2 \simeq 10^{-2}$. It should be noted that these suppression factors are not general and do not always appear \cite{Jenkins:2013fya}. For instance, compensation factors might arise if the fundamental theory is strongly coupled. In case such loop factors do appear, the bounds on the quark (C)EDMs are, even at the relatively low scale $\MT=1\,\mathrm{TeV}$, not particularly strong. The couplings $\Sigma_{1,8}$, $\Xi_{1}$, and $Y^{\prime u,d}$ can be generated at tree level but they might still be suppressed by small couplings or phases. For example, in the minimal left-right symmetric model, $Y^{\prime u,d}$ is suppressed by the SM light-quark Yukawa couplings \cite{Zhang:2007da}.

\begin{table}[t]
\caption{\small Bounds on the coupling constants of various dimension-six operators. All entries are dimensionless.}
\begin{center}\footnotesize
\begin{tabular}{||c|cc||}
\hline
& $\MT = 1\,\mathrm{TeV}$&$\MT=10\,\text{TeV}$ \\
\hline 
  $(\MT^2)d_{u,\,d}\left(\MT\right)$ & $\leq \lbrace 1.8,\,1.8 \rbrace\cdot 10^{-3}$ & $\leq \lbrace 2.1,\,2.1 \rbrace\cdot 10^{-1}$\\
  $(\MT^2)\tilde{d}_{u,\,d}\left(\MT\right) $& $\leq \lbrace 1.9,\,0.91 \rbrace\cdot 10^{-3}$&  $\leq \lbrace 1.7,\,0.94 \rbrace\cdot 10^{-1}$\\
  $(\MT^2)d_W\left(\MT\right) $&$\leq 5.6\cdot 10^{-5}$& $\leq 7.0\cdot 10^{-3}$\\
  $(\MT^2)\mathrm{Im}\,\Sigma_1 \left(\MT\right)$ &$\leq 3.2\cdot 10^{-5}$ &$\leq 2.3\cdot 10^{-3}$\\
  $(\MT^2) \mathrm{Im}\,\Sigma_8\left(\MT\right)$ &$\leq 3.3\cdot 10^{-4}$ &$\leq 2.4\cdot 10^{-2}$\\
  $(\MT^2)\mathrm{Im}\,\Xi_{1} \left(\MT\right)$ &$\leq 1.7\cdot 10^{-4}$ &$\leq 1.7\cdot 10^{-2}$\\
  $(\MT^2) \mathrm{Im}\,Y^{\prime u,\,d}\left(\MT\right)$&$\leq \lbrace{8.9,\,8.9\rbrace}\cdot 10^{-5}$&$ \leq \lbrace{7.9,\,7.9\rbrace}\cdot 10^{-3}$\\
$ (\MT^2) \theta^\prime\left(\MT\right)$&$\leq 2.4\cdot 10^{-3}$&$ \leq 1.5\cdot 10^{-1}$\\
  \hline
\end{tabular}
\end{center}
\label{TableMTMQCDBounds} 
\end{table}

\section{Suppressed operators}\label{sup}
In general, we expect the operators in Eq.\ \eqref{set1MQCD} (which descend from Eq.\ \eqref{set1a}) to describe the dominant part of $P$ and $T$ violation due to physics beyond the SM in first-generation hadronic and nuclear systems. However,  in specific models of new physics the associated coupling constants could be much smaller, or even zero, due to additional symmetry considerations or unexpected fine-tuning. In such models the most important operators might be operators which were neglected so far such as the dimension-six operators mentioned in Sec.\ \ref{operators} that obtain additional suppression at low energies. Higher-dimensional operators might also become relevant, but we limit the discussion to the dimension-six operators. We assume that for some reason the operators of the last section are suppressed, thereby leaving room for the remaining dimension-six operators. In order to keep the discussion organized, we first discuss the $\slashPT$ quark-Higgs operators in Sec.\ \ref{supqqH}. These operators give rise to a rich set of $\slashPT$ four-quark operators at low energies. In Sec.\ \ref{heavyboson} we discuss the remaining $\slashPT$ operators which involve heavy gauge bosons.

\subsection{$P$- and $T$-violating quark-Higgs interactions}\label{supqqH}
We begin the analysis with Eq.\ \eqref{qqH} describing interactions between light quarks and Higgs bosons
\begin{eqnarray} \label{qqH2}
\vL_{6,qq\vp\vp\vp}&=&h\left(v+h\right)\left(v+\frac{h}{2}\right)\left(
 i\, \mathrm{Im}\, Y^{\prime u}\,\bar u \g^5 u + i\, \mathrm{Im}\, Y^{\prime d}\,\bar d \g^5 d \right)\nn\\
 &=& \sum_{q=u,d} C_{qH}\,O_{qH} + \dots,
\end{eqnarray}
where the dots denote terms with more than one Higgs boson and Eqs.\ \eqref{set1a} and \eqref{MT} were applied. These interactions have already been considered in the previous section where they induced relatively large corrections to the quark CEDMs via two-loop diagrams involving top quarks \cite{Barr:1990vd}. Here we assume that these corrections, together with the other contributions to the quark (C)EDMs, are suppressed. This requires fine-tuning between different coupling constants which is not expected on general grounds but might occur in specific models of new physics. 

Starting at the high-energy scale $\MT$, the Higgs-quark interactions are first evolved down to $\MW = M_W$. The anomalous dimension is given in Eq.\ \eqref{gammaqqH} and the solution of the RGE is 
\bea \label{qqHMTMW}
C_{qH}(M_W) = \left(\frac{\al_s(M_W)}{\al_s(m_t)}\right)^{12/23}\left(\frac{\al_s(m_t)}{\al_s(\MT)}\right)^{4/7} \,C_{qH}(\MT).
\eea
Numerically this becomes for two explicit values of $\MT$
\bea \label{qqHnumeric}
C_{qH}(M_W) = 1.2\, C_{qH}(1 \,\text{TeV}),\qquad C_{qH}(M_W) = 1.3\, C_{qH}(10 \,\text{TeV}). 
\eea
Below $M_W$ the Higgs boson is integrated out which at tree-level generates $\slashPT$ four-quark operators among light quarks of the following form
\bea\label{4qsuppressed}
\vL_{6,\, qqqq} &=& -\frac{iv}{M_H^2}\bigg(m_u C_{uH}(M_W)\,\bar u\g^5u\,\bar u u + m_d C_{dH}(M_W)\,\bar d\g^5d\,\bar d d \nn\\
&&+m_dC_{uH}(M_W)\,\bar u\g^5u\,\bar d d + m_uC_{dH}(M_W)\,\bar u u\,\bar d \g^5 d\bigg).
\eea
Since $C_{qH} = \Or(1/M_{\slashTsub}^2)$ these interactions scale as $\Or(v m_q/M_{H}^2 M_{\slashTsub}^2) \simeq \Or(m_q/v M_{\slashTsub}^2)$ such that they are not only suppressed by two powers of $M_{\slashTsub}$ but also by the ratio of the light-quark mass to the Higgs vev, $v\simeq 246\,\mathrm{GeV}$. The structure of the four-quark operators appearing in Eq.\ \eqref{4qsuppressed} is different from that in the previous section, which calls for an extension of the operator basis in Eqs.\ \eqref{set1a} and \eqref{basisextended} with six additional four-quark operators
\bea\label{set2}
\tilde O_{PS_1} &=& \frac{ig^2_s}{2} (\bar u u \hspace{1mm} \bar d \g_5 d+\bar u \g_5 u \hspace{1mm} \bar d d +\bar d u \hspace{1mm} \bar u \g_5 d+\bar d \g_5 u \hspace{1mm} \bar u d),\nn\\
\tilde O_{PS_8} &=& \frac{ig^2_s}{2} (\bar u t^a u \, \bar d \g_5 t^a d+\bar u \g_5 t^a u \, \bar d t^a d + \bar d t^a u \, \bar u \g_5 t^a  d + \bar d \g_5 t^a u \, \bar u t^a d),\nn\\
O_{4q_1} &=&  i \, g^2_s\,(\bar q \g^5 q\,\bar q q),\nn\\
O_{4q_8} &=&  i \, g^2_s\,(\bar q \g^5 t^a q\,\bar q t^a q),
\eea
where again $q \in \{u , d\}$. Using the Fierz identities 
\begin{eqnarray}
i g^2_s\,\bar u \g^5 u\,\bar d d &=& \frac{1}{2}(O_{PS_1} + \tilde O_{PS_1})+\frac{1}{6}O_{LR_1} + O_{LR_8},\nn\\
i g^2_s\,\bar u  u\,\bar d \g^5 d &=& \frac{1}{2}(O_{PS_1}+ \tilde O_{PS_1})-\frac{1}{6}O_{LR_1} - O_{LR_8},
\end{eqnarray}
the coupling constants $C_i(\mu)$ at the scale $M_W^-$ are found to be 
\bea\label{MWminSup}
C_{PS_{1}}(M_W^-)\!\! &=&\tilde C_{PS_{1}}(M_W^-)\,\,=\,\,-\frac{1}{2}\frac{1}{g^2_s(M_W^+)}\left(\frac{m_dv}{M_H^2} C_{uH}(M_W^+) + \frac{m_u v}{M_H^2}C_{dH}(M_W^+)\right),\nn\\
C_{LR_{1}}(M_W^-)\!\! &=&\frac{1}{6} C_{LR_{8}}(M_W^-)\,\,=\,\,- \frac{1}{6}\frac{1}{g^2_s(M_W^+)}\left(\frac{m_d v}{M_H^2}C_{uH}(M_W^+) -\frac{m_u v}{M_H^2} C_{dH}(M_W^+)\right),\nn\\
C_{4q_1}(M_W^-)\!\! &=&\!\! -\frac{1}{g^2_s(M_W^+)} \frac{m_q v}{M_H^2}C_{qH}(M_W^+) ,\nn\\
C_{PS_{8}}(M_W^-)\!\! &=& \tilde C_{PS_{8}}(M_W^-) \,\,=\,\, C_{4q_8}(M_W^-)\,\,=\,\ 0.
\eea

\begin{figure}[t]
\centering
\includegraphics[scale = 0.7]{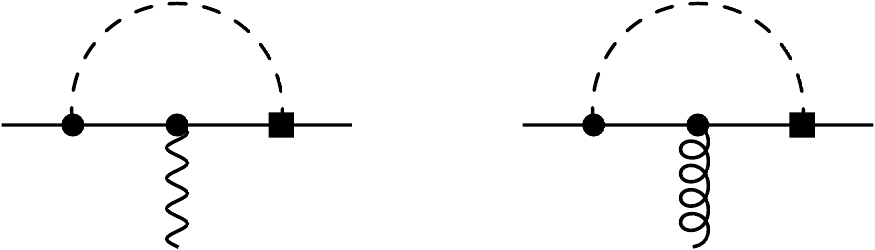}
\caption{One-loop diagrams contributing to the quark electric and chromo-electric dipole moments.
A square marks a $\slashPT$ interaction from Eq.\ \eqref{qqH2}. Other notation is as in Figs. \ref{FQto(C)EDM} and \ref{GluonHiggstoCEDM}. For simplicity only one possible ordering is shown here.}
\label{HiggsToEDM}
\end{figure}

The $\slashPT$ quark-Higgs interactions induce the quark EDM and CEDM through one-loop diagrams shown in Fig. \ref{HiggsToEDM}. These diagrams are finite and give the following contributions \cite{Hisano:2012cc}
\bea\label{HiggsLoop}
C_q(M_W^-) = -\tilde C_q(M_W^-) =  \frac{1}{(2\pi)^2} \frac{m_qv}{M_H^2}\left(\frac{3}{4}+\ln \frac{m_q}{m_H}\right)C_{qH}(M_W^+),
\eea
which are smaller by a factor $\Or(10^{-5})$ compared to Eq.\ \eqref{qqHunsup}. This is approximately the level of fine-tuning needed in order to make the four-quark operators in Eq.\ \eqref{4qsuppressed} significant at low energies. Eqs.\ \eqref{MWminSup} and \eqref{HiggsLoop} represent matching conditions at $\mu = M_W$, meaning that the quark masses appearing there should be evaluated at this scale as well. The one-loop QCD running gives  $m_q (M_W) \simeq 0.58\, m_q(\MQCD)$. In all results below we will use $m_q$ to denote the quark mass at $\MQCD$.

The next step is to evolve the operators down to $\MQCD$. We choose the basis 
\bea \label{basissup}
\vec C (\mu)&=& (C_{q}(\mu), \tilde C_{q}(\mu), C_W(\mu),  C_{PS_1}(\mu), C_{PS_8}(\mu), C_{LR_1}(\mu), C_{LR_8}(\mu),\nonumber\\
&& \tilde C_{PS_1}(\mu), \tilde C_{PS_8}(\mu), C_{4q_1}(\mu), C_{4q_8}(\mu))^T ,
\eea
with the extended RGE 
\bea \label{RGE2Ext}
\frac{d \vec C(\mu)}{d \ln \mu} = \g \,\vec C(\mu).
\eea
Up to first order in $\alpha_s$, $\g$ can be written as 
\bea \label{gamma2}
\ga = \frac{\al_s}{4\pi} \bma \gamma_{\mathrm{dipole}} & \gamma_{\mathrm{mix}}  & 0&\tilde \gamma_{\mathrm{mix}} & \gamma_{4q,\mathrm{mix}} \\
0&\gamma_{PS} &0 &0 &0\\
0 &0 & \gamma_{LR}&0 &0\\
0 & 0 & 0 & \tilde \gamma_{PS} & 0\\
0 & 0 & 0 & 0 & \gamma_{4q}\ema.
\eea

The top-left entries of this matrix are given in Eqs.\ \eqref{gammadipole}, \eqref{gamma4quark}, \eqref{gammamix}, and \eqref{gammaFQLR}. The other entries denote the running and mixing of the four-quark operators in Eq.\ \eqref{set2}. As is already clear from the form of $\gamma$, these additional four-quark operators do not mix with the FQPS and FQLR operators. Furthermore, $\tilde O_{PS,1,8}$ do not mix with $O_{4q,1,8}$ at the one-loop level.
These considerations justify the block-diagonal form of $\g$, apart from $\gamma_{\mathrm{mix}}$, $\tilde \gamma_{\mathrm{mix}}$, and $\gamma_{4q,\mathrm{mix}}$ which give rise to contributions to the quark (C)EDMs from the various four-quark interactions. Explicit calculation shows that $\tilde \gamma_{PS}$ is identical to $\gamma_{4q}$ 
\bea 
\tilde \g_{PS}  = \gamma_{4q} =  2\bma 
\frac{(4-3N)(N\sq-1)}{N\sq}+\bt_0 & \frac{(N-1)\sq(1+N)(2+N)}{N^3} \\ 
4\frac{N-2}{N} & 2C_2(N) + 2\frac{N\sq+2}{N\sq}+\bt_0
\ema. 
\eea 
Apart from a sign $\tilde O_{PS,1,8}$ mixes into the quark (C)EDMs the same way as $O_{PS,1,8}$ such that
$\tilde \gamma_{\mathrm{mix}}= - \gamma_{\mathrm{mix}}$. 
The four-quark operators containing a single flavor contribute to the dipoles via
\bea
\gamma_{4q,\mathrm{mix}} = 4 \bma -1 & -C_2(N) \\
1& C_2(N) - \frac{N}{2}\\
0&0 \ema 
,\eea
where it is implied that the operator containing only up (down) quarks induces only the up (down) quark (C)EDM. The results for the anomalous dimension matrices agree with those in Refs. \cite{An:2009zh, Hisano:2012cc} after a basis transformation.

Following the same procedure as in Sec.\ \ref{sec32} we now find, apart from Table \ref{TableMWMQCD}, additional contributions, shown in Table \ref{TableMWMQCDSup}, to the quark (C)EDMs. The results for the gluon CEDM, the FQPS, and the FQLR operators are unchanged.
\begin{table}
\caption{Dependence of the quark electric and chromo-electric dipole moments and four-quark operators at $\MQCD$ on the same four-quark operators at $M_W$. }
\begin{center}
\begin{tabular}{||c|cccc||}
\hline 
\rule{0pt}{2.5ex}  & $\tilde C_{PS_1}(M_W)$  & $\tilde C_{PS_8}(M_W)$&$C_{4q_1}(M_W)$& $C_{4q_8}(M_W)$  \tabularnewline
\hline 
$C_q(\MQCD)$& $\left(0.070 \frac{Q_{q^\prime}}{Q_q}-0.027\right) \frac{m_{q^\prime}}{m_q}$&$ \left(0.045 \frac{Q_{q^\prime}}{Q_q}+0.0058\right) \frac{m_{q^\prime}}{m_q}$& $0.085$ & $0.10$\tabularnewline
$\tilde C_q (\MQCD)$ & $-0.080\frac{m_{q^\prime}}{m_q}$ &$0.018\frac{m_{q^\prime}}{m_q}$&$-0.16$&$0.037$  \tabularnewline
$\tilde C_{PS_1}(\MQCD)$ &$0.63$&$-0.14$&$-$&$-$\tabularnewline
$\tilde C_{PS_8}(\MQCD)$&$-0.063$&$0.18$&$-$&$-$ \tabularnewline
$C_{4q_1}(\MQCD)$&$-$&$-$&$0.63$&$-0.14$ \tabularnewline
$C_{4q_8}(\MQCD)$&$-$&$-$&$-0.063$&$0.18$\tabularnewline
\hline
\end{tabular}
\end{center}
\label{TableMWMQCDSup} 
\end{table}
The following results for the quark (C)EDMs are obtained for $\MT = 1 \, \text{TeV}$
\bea \label{q(C)EDMlowSup1}
d_q(M_{\text{QCD}})& = &
\left[0.0020\left(0.21+\ln\frac{m_q}{M_H}\right)-0.038\right]\frac{m_qv}{M_H\sq}\mathrm{Im}\,Y^{\prime q}(1 \, \text{TeV})\nn\\
&&+\frac{m_{q'}}{m_q}\left[0.00035\frac{Q_{q'}}{Q_q}-0.0028\right]\left(\frac{m_u v}{M_H\sq}\mathrm{Im}\,Y^{\prime d}(1 \, \text{TeV})+\frac{m_d v}{M_H\sq}\mathrm{Im}\,Y^{\prime u}(1 \, \text{TeV})\right),\nn\\
\tilde d_q(M_{\text{QCD}})& = &
\left[0.15-0.020\left(0.21+\ln\frac{m_q}{M_H}\right)\right]\frac{m_qv}{M_H\sq}\mathrm{Im}\,Y^{\prime q}(1 \, \text{TeV})\nn\\
&&-0.028\frac{m_{q'}}{m_q}\left(\frac{m_u v}{M_H\sq}\mathrm{Im}\,Y^{\prime d}(1 \, \text{TeV})+\frac{m_d v}{M_H\sq}\mathrm{Im}\,Y^{\prime u}(1 \, \text{TeV})\right),
\eea
while for $\MT = 10 \, \text{TeV}$ the results are trivially obtained from Eq.\ \eqref{qqHnumeric}.
Because the logarithm is rather large, $\ln m_q/M_H \simeq -10$, the one-loop diagrams contribute at the same order as the mixing of the four-quark operators into the dipoles. However, all terms receive a suppression of $m_q/v$ compared to Eqs.\ \eqref{q(C)EDMlow1} and \eqref{q(C)EDMlow10}. 

The contributions to the various four-quark operators are suppressed by $m_q/v$ as well. For the four-quark operators appearing in Sec.\ \ref{unsup}, we find
\bea\label{FQPSsup}
\lbrace \mathrm{Im}\,\Sigma_1(M_{\text{QCD}}),\, \mathrm{Im}\,\Sigma_8(M_{\text{QCD}}) \rbrace =\lbrace -1.6,\,0.73 \rbrace \left[\frac{m_d v}{M_H\sq}\mathrm{Im}\,Y^{\prime u}( 1\,\text{TeV})+\frac{m_u v}{M_H\sq}\mathrm{Im}\,Y^{\prime d}( 1\,\text{TeV})\right],\nn\\
\lbrace \mathrm{Im}\,\Xi_{LR_1}(M_{\text{QCD}}),\, \mathrm{Im}\,\Xi_{LR_8}(M_{\text{QCD}}) \rbrace =-\lbrace 0.34,\,2.0 \rbrace \left[\frac{m_d v}{M_H\sq}\mathrm{Im}\,Y^{\prime u}( 1\,\text{TeV})-\frac{m_u v}{M_H\sq}\mathrm{Im}\,Y^{\prime d}( 1\,\text{TeV})\right].
\eea
A comparison of Eqs.\ \eqref{q(C)EDMlowSup1} and \eqref{FQPSsup} makes it clear that  the four-quark operators are larger by one to two orders of magnitude than the quark (C)EDMs. In the specific scenario discussed here, it seems safe to neglect the latter.

Finally, there are the remaining six four-quark operators in Eq.\ (\ref{set2}).
In order to compare with Eq.\ \eqref{FQPSsup} it is useful to redefine these operators without the factor $g_s^2$ and define new coupling constants 
\bea\label{set2a}
\tilde \Sigma_{1,8}(\mu) &\equiv & g_s^2(\mu)\,\tilde{C}_{\text{PS}_{1,8}}(\mu),\nn\\
\Omega_{q,1}  (\mu) &\equiv & g_s^2(\mu)\, C_{4q_1}(\mu),\nn\\
 \Omega_{q,8}  (\mu) &\equiv & g_s^2(\mu)\,C_{4q_8}(\mu),
\eea
for which we find
\bea\label{redef}
\lbrace \tilde \Sigma_{1}\left(M_{\text{QCD}}\right),\,\tilde \Sigma_{8}\left(M_{\text{QCD}}\right) \rbrace &=&  \lbrace -0.64,\,0.064 \rbrace \left[\frac{m_d v}{M_H\sq}\mathrm{Im}\,Y^{\prime u}(1\,\text{TeV})+\frac{m_u v}{M_H\sq}\mathrm{Im}\,Y^{\prime d}(1\,\text{TeV})\right],\nn\\
\lbrace  \Omega_{q,1}\left(M_{\text{QCD}}\right),\,\Omega_{q,8}\left(M_{\text{QCD}}\right) \rbrace &=&  \lbrace -1.3,\,0.13 \rbrace\, \frac{m_qv}{M_H\sq} \mathrm{Im}\,Y^{\prime q}\left(1\,\text{TeV}\right).
\eea
These results indicate that the color-singlet operators with couplings $\tilde  \Sigma_{1}$ and $\Omega_{q_1}$, are larger by an order of magnitude than their color-octet counterparts $\tilde  \Sigma_{8}$ and $\Omega_{q_8}$. Thus, at low energies, it is sufficient to consider only the color-singlet operators which are of  the same order as $\mathrm{Im}\,\Sigma_{1,8}$ and $\mathrm{Im}\,\Xi_{LR_{1,8}}$.

To summarize this section we write the $\slashPT$ Lagrangian at the scale $M_{\text{QCD}}$ in case the dominant dimension-six operators at the electroweak scale are the Higgs-quark interactions in Eq.\ \eqref{qqH2}. Neglecting, as discussed above, the quark (C)EDMs and the $\tilde  \Sigma_{8}$ and $\Omega_{q,8}$ four-quark operators, the Lagrangian at $\MQCD$ is given by
\bea \label{set2MQCD}
\vL_{6, \mathrm{set}\,2} &=&  +i \frac{\mathrm{Im}\, \Sigma_{1}}{2} (\bar u u \, \bar d \g_5 d+\bar u \g_5 u \, \bar d d -\bar d u \, \bar u \g_5 d-\bar d \g_5 u \, \bar u d)\nonumber\\
&&+ i \frac{\mathrm{Im}\, \Sigma_{8}}{2}  (\bar u t^a u \, \bar d \g_5 t^a d+\bar u \g_5 t^a u \, \bar d t^a d -\bar d t^a u \, \bar u \g_5 t^a  d-\bar d \g_5 t^a u \, \bar u t^a d)\nonumber\\
&& + i\, \mathrm{Im}\,\Xi_{LR_1} \left(\bar u_R \g^\mu d_R\,\bar d_L \g_\mu u_L - \bar d_R \g^\mu  u_R\,\bar u_L \g_\mu d_L \right) \nn\\
&& + i\, \mathrm{Im}\,\Xi_{LR_8} \left(\bar u_R \g^\mu t^a d_R\,\bar d_L \g_\mu t^a u_L - \bar d_R \g^\mu t^a u_R\,\bar u_L \g_\mu t^a d_L \right)  \nn\\
&&+i \frac{\tilde \Sigma_{1}}{2} (\bar u u \hspace{1mm} \bar d \g_5 d+\bar u \g_5 u \hspace{1mm} \bar d d +\bar d u \hspace{1mm} \bar u \g_5 d+\bar d \g_5 u \hspace{1mm} \bar u d)\nn\\
&& +  i\,\Omega_{u,1} (\bar u \g^5 u\,\bar u u) + i  \,\Omega_{d,1}( \bar d \g^5 d\,\bar d d),
\eea
where the coupling constants are all evaluated at $\MQCD$ and their values in term of the coupling constants in Eq.\ \eqref{qqH2} are given in Eqs.\ \eqref{FQPSsup} and \eqref{redef}. All coupling constants in Eq.\ \eqref{set2MQCD} scale as $(m_q/v)(1/ M^2_{\slashTsub})$ and, in general, are much smaller than the constants appearing in Eq.\ \eqref{set1MQCD}. We stress again that the new four-quark operators appearing in this section are only important in  specific scenarios involving additional symmetries and/or fine-tuning. In the scenario sketched here there are seven relevant $\slashPT$ four-quark operators around $\MQCD$ with only two independent coupling constants. 

\subsection{$P$ and $T$ violation involving heavy gauge bosons}\label{heavyboson}
In this section we deal with the remaining operators in Sec.\ \ref{operators} which contain heavy $W^{\pm}$-, $Z$-, and Higgs-boson fields. These operators  will no longer exist in the effective theory below the electroweak scale, but they will contribute to operators containing light fields only.  
The operators in this section only give rise to low-energy $PT$ violation via electroweak one-loop diagrams. The operators studied are the weak dipole moments of the quarks in Eqs.\ \eqref{qZEDMdef} and \eqref{qWEDMdef} and the interactions among electroweak gauge and Higgs bosons in Eqs.\ \eqref{Wweakdipole}, \eqref{dim6thetahiggs}, and \eqref{thetawb}.
 
These operators contribute to quark (C)EDMs through various diagrams, however, in all cases the contributions are suppressed by electroweak coupling constants in the typical combination $\alpha_w = e^2/4\pi$. The quark (C)EDMs are generated when the effective operators are evolved from $\MT$ to $\MW$. Below $\MW$ the heavy bosons can be integrated out and the running of the quark (C)EDMs to $\MQCD$ can simply be obtained from the results in Sec.\ \ref{unsup}. 
In principle, the quark (C)EDMs originate not only from the running from $M_{\slashTsub}$ to $\MW$, but also from threshold corrections at $M_W$, \textit{i.e.} the finite parts of the Feynman diagrams. Here we do not calculate these parts unless the operator in question does not produce a quark (C)EDM otherwise. This is only necessary for the interactions among gauge bosons with coupling constant $d_w$ in Eq.\ \eqref{Wweakdipole}. For the other operators we assume the sizes of the induced quark (C)EDMs to be saturated by the running part. We do not expect this approximation to significantly alter the results.

\begin{figure}[t!]
\centering
\includegraphics[scale = 0.7]{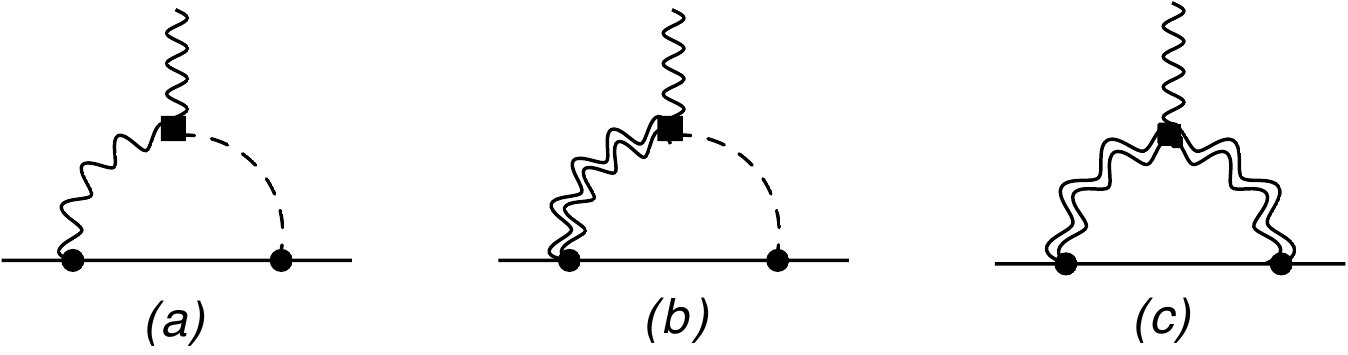}
\caption{One-loop diagrams contributing to the quark electric moment. The double wavy lines denote the propagation of  $W^{\pm}$ and $Z$ bosons. The black square denotes a vertex from one of the operators in Eq.\ \eqref{HeavyBosonOperators}. The other notation is as in Figs. \ref{FQto(C)EDM} and \ref{GluonHiggstoCEDM}. For simplicity only one possible ordering is shown here.}
\label{HiggsGaugeToEDM}
\end{figure}
To calculate the running of the operators we will take the $\Or(\al_w)$ contributions as a perturbation to the QCD RGE. We only consider the electroweak corrections which induce a quark (C)EDM. That is, we do not consider the electroweak running of $\al_w$ itself, nor that of the operators involving the heavy bosons. These effects would give rise to $\Or(\al_w^2)$ corrections to the induced quark (C)EDMs.
The RGE is written as
\bea \label{HeavyRGE}
\frac{\partial \vec C}{\partial \ln \mu} = \ga \, \vec C , \hspace*{5mm}
\vec C = (C_q,\tilde C_q , C_H)^T,
\eea
where $C_H$ stands for the coupling constant of one of the heavy-boson operators. These will be treated one at a time because their mixing can be neglected at the order we work. 
We write $\vec C = \vec C^{(0)}+ \vec C^{(1)}$ and $\ga = \ga^{(0)} + \ga^{(1)}$ where the superscript $0$ $(1)$ denotes terms of $\Or(\al^0_w)$ ($\Or(\al_w^1)$). Up to $\Or(\al_w^1)$, Eq.\ \eqref{HeavyRGE} splits into two equations
\bea \label{HeavyBosonRGE}
\frac{\partial \vec C^{(0)}}{\partial \ln \mu} = \ga^{(0)} \, \vec C^{(0)} , \qquad
\frac{\partial \vec C^{(1)}}{\partial \ln \mu} = \ga^{(0)} \, \vec C^{(1)} +\ga^{(1)} \, \vec C^{(0)} ,
\eea
where $\ga^{(0)}$ describes the effects of QCD corrections, while $\ga^{(1)}$ describes the mixing of the operators due to electroweak corrections. The LO matrix can, for our purposes, be written as
\bea
\ga^{(0)} = \frac{\al_s}{4\pi} \bma \ga_e & \ga_{qe} & 0\\0 & \ga_q & 0\\ 0&0&\ga_H \ema ,
\eea
where $\ga_{e,q,qe}$ can be read off from Eq.\ \eqref{gammadipole} ($\g_e=-\g_{qe}=8 C_2(N)$ and $\g_q = 16C_2(N)-4N$) while $\ga_H$ and the matrix $\ga^{(1)}$ depend on the heavy-boson operator under investigation. The electroweak effects can, at the order we work, be described by
\bea
\ga^{(1)} = \frac{\al_w}{4\pi} \bma 0 & 0 & \ga^{(1)}_{13}\\0 & 0&  \ga^{(1)}_{23}\\ 0&0&0 \ema.
\eea
What remains is the calculation of the unknown entries $\ga_H$, $\ga^{(1)}_{13}$, and $\ga^{(1)}_{23}$ for each operator.

We begin by defining the relevant operators,
\bea \label{HeavyBosonOperators}
O_{B} &=&  e\sq \varepsilon^{\mu\nu\al\bt} \left(F_{\mu\nu} F_{\al\bt}-2 \frac{s_w}{c_w}F_{\mu\nu} Z_{\al\bt} \right)\,vh+\dots,\nn\\
O_{W} &=& e\sq \varepsilon^{\mu\nu\al\bt} \left(F_{\mu\nu} F_{\al\bt}+2 \frac{c_w}{s_w}F_{\mu\nu} Z_{\al\bt} \right)\,vh+\dots,\nn\\
O_{WB} &=&e\sq v^2 \varepsilon^{\mu\nu\al\bt}\left\{\left[\Fmu F_{\al\bt} +\left(\frac{c_w}{s_w} - \frac{s_w}{c_w}\right)\Fmu Z_{\al\bt}\right]\frac{h}{v}- i\frac{g}{s_w}\,  W_\mu^+ W_\nu^- F_{\al\bt} \right\}+\dots,\nn\\
O_{d_W} &=& i \varepsilon^{\mu\nu\al\bt} W_{\beta}^{+\,\rho} W_{\rho\al}^{-}\Fmu+\dots,\nn\\
O_{Wu} &=&-\frac{g}{\sqrt{2}} m_u\left[(\bar d_L i \simu  u_R )\,W^-_{\mu\nu}-i \frac{g}{\sqrt{2}} (\bar u i \simu \g_5 u )\,W_\mu^+ W_\nu^- + \text{h.c.}\right],\nn\\
O_{Wd} &=&-\frac{g}{\sqrt{2}} m_d\left[(\bar u_L i \simu  d_R)\, W^+_{\mu\nu}+i \frac{g}{\sqrt{2}} (\bar d i \simu \g_5 d )\,W_\mu^+ W_\nu^- + \text{h.c.}\right],\nn\\
O_{Zq} &=& -\frac{g}{2}m_q \,\bar q i\simu \g_5 q\, Z_{\mu\nu},
\eea
where the dots denote terms which are not relevant at $\Or(\al_w)$. The coupling constants at the scale $\MT$ are
\bea
C_{B}(\MT)&=& \theta_B^\prime(\MT),\qquad\qquad
C_{W}(\MT)= \theta_W^\prime(\MT),\nn\\
C_{WB}(\MT)&=&\theta_{WB}^\prime(\MT),\qquad\hspace{3mm}
C_{d_W}(\MT)=-s_wd_w(\MT),\nn\\
C_{Wq}(\MT) &=& w_q(\MT),\qquad\qquad C_{Zq}(\MT) = z_q(\MT).
\eea

\begin{table}[b]
\caption{\small The anomalous dimensions of the various heavy boson operators. Here $T^3_q$ stands for the third component of weak isospin of the external quark, \textit{i.e.} $1/2$ $(-1/2)$ for the up (down) quark.}
\begin{center}\small
\begin{tabular}{||c|ccc||}
\hline 
 & $\g_{13}^{(1)}$ & $\g_{23}^{(1)} $& $\g_{H}  $ \\
 \hline
 $O_B$ & $-16\left(1-\frac{T^3_q /2-s_w\sq Q_q}{Q_q c_w^2}\right)$ & 0 & 0\\
 $O_W$ & $-16\left(1+\frac{T^3_q /2-s_w\sq Q_q}{Q_q s_w\sq }\right)$ & 0 &0\\
 $O_{WB}$ & $-4\frac{1}{s_w\sq  }\left(2T^3_q V_{ud}\sq\frac{1}{Q_q}+4s_w\sq +\frac{T_q^3-2s_w\sq Q_q}{Q_q  c_w\sq}\cos 2\theta_w \right)$ & 0 &0\\
 $O_{d_W}$ & 0 & 0 &0\\
 $O_{Wq}$ &$ \frac{1}{s_w\sq }\left[V_{ud}\left(5-9\frac{Q_{q'}}{Q_q}
\right)-6\left(1-\frac{Q_{q'}}{Q_q}
\right) \right]$ & $ \frac{4}{ s_w\sq }V_{ud}$ & $8 C_2(N)$\\
 $O_{Zq}$ &$ -4\frac{T^3_q-2 Q_q s_w\sq}{c_w s_w\sq}$ & $4\frac{T^3_q-2 Q_q s_w\sq}{c_w s_w\sq}$& $8 C_2(N) $\\
  \hline
\end{tabular}
\end{center}
\label{TableGammas} 
\end{table}
The anomalous dimensions for the heavy boson operators are collected in Table \ref{TableGammas}. The electroweak loops for $O_{WB}$, $O_{d_W}$, $O_{Zq}$, and $O_{Wq}$ were evaluated in Ref. \cite{DeRujula:1990db} in a different regularization scheme. However, for $O_{Wq}$ not all diagrams were taken into account. 

The operators $O_B$, $O_W$, $O_{WB}$, and $O_{d_W}$ induce a quark EDM through the diagrams in Fig. \ref{HiggsGaugeToEDM}. The first two operators contribute through Diagrams \ref{HiggsGaugeToEDM}(a,b) and $O_{d_W}$ through Diagram \ref{HiggsGaugeToEDM}(c). $O_{WB}$ contributes via all diagrams. None of these operators induces a quark CEDM at the one-loop level.
\begin{figure}[t]
\centering
\includegraphics[scale = 0.7]{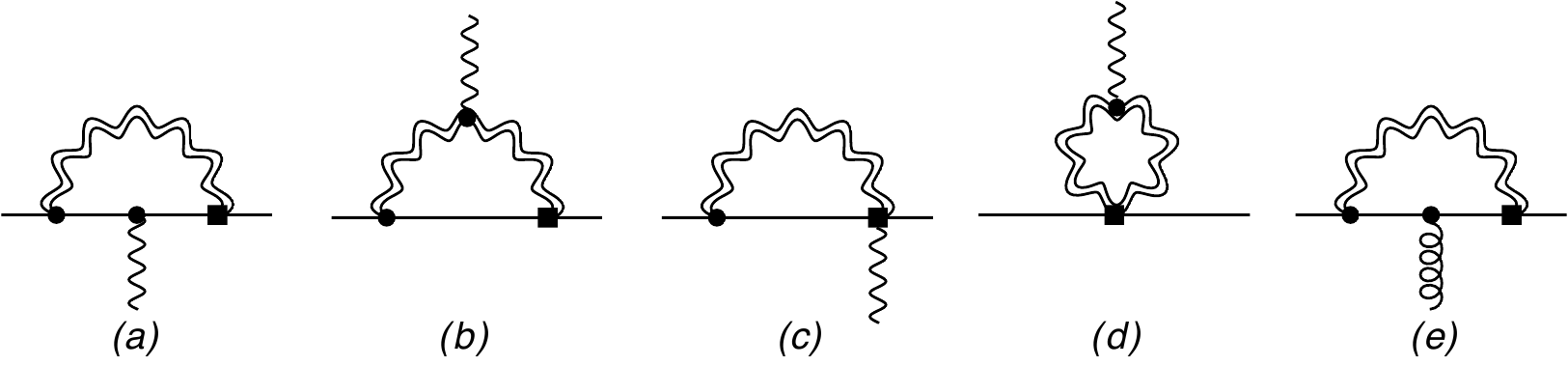}
\caption{One-loop diagrams contributing to the quark electric and chromo-electric dipole moments. The notation is as in Fig. \ref{HiggsGaugeToEDM}. For simplicity only one possible ordering is shown here.}
\label{HeavyToEDM}
\end{figure}
As can be seen from Table \ref{TableGammas}, the anomalous dimensions of $O_{d_W}$ vanish. The reason is that for this operator Diagram \ref{HiggsGaugeToEDM}(c) is finite in dimensional regularization. However, the diagram does have a finite part which was evaluated in several regularization schemes in Ref. \cite{Boudjema:1990dv}, where it was found to be scheme dependent. We will use the result found in dimensional regularization. In our notation the operator $O_{d_W}$ induces the quark EDM \cite{Boudjema:1990dv}
\bea
C_{u,d}(M_W^-) = \pm\frac{g\sq}{32\pi\sq}\frac{1}{e Q_{u,d}} C_{d_W},
\eea
where the plus (minus) sign is for the up (down) quark EDM.

The weak dipole moments, $O_{Zq}$ and $O_{Wq}$, contribute to four-quark operators at tree level. However, this necessarily involves one power of the exchanged momentum because of the derivative acting on the gauge fields. The coupling constants of these effective dimension-seven operators then scale as $m_q/M_W^2\,1/M_{\slashTsub}^2$ and are heavily suppressed. The same holds for one-loop contributions to four-quark operators. Larger effects come from one-loop diagrams shown in Fig. \ref{HeavyToEDM}(a-d) and \ref{HeavyToEDM}(e) contributing to, respectively, light-quark EDMs and CEDMs. The weak dipole moments are the only operators in this section which produce a quark CEDM. Furthermore, they are the only operators which are affected by QCD corrections and, hence, have a nonzero $\gamma_H$.

\begin{table}[t]
\caption{\small The contributions of the various heavy boson operators at $\MT$ to the q(C)EDMs at $\MQCD$ in units of $10^{-2}$. A ``$-$" indicates that there is no contribution at this order.}
\begin{center}\footnotesize
\begin{tabular}{||c|cccccc||}
\hline 
 & $C_B\left(1\, \text{TeV}\right)$ & $C_W\left(1\, \text{TeV}\right) $& $C_{WB}\left(1\, \text{TeV}\right) $&$ C_{d_W}(1\, \text{TeV})$ &$ C_{Wq}(1\, \text{TeV})$&$ C_{Zq}(1\, \text{TeV})$ \\
 \hline
 $C_{ u,\,d}\left(M_W\right)$ &$ \lbrace 1.8,\, 0.73\rbrace$ & $\lbrace3.6,\,7.3\rbrace $&$ \lbrace 6.2,\,11\rbrace $& $\lbrace  0.65,\,1.3\rbrace$ &$ \lbrace -13,\,-15\rbrace$&$ \lbrace -2.4,\,4.3\rbrace$\\
$ \tilde{C}_{ u,\,d}\left(M_W\right) $& $-$ & $-$ &$-$&$-$ &$  \lbrace-2.2, -2.2 \rbrace$ &$ \lbrace-0.49,\, 0.88\rbrace$\\
\hline 
 & $C_B\left(10\, \text{TeV}\right)$ & $C_W\left(10\, \text{TeV}\right) $& $C_{WB}\left(10\, \text{TeV}\right) $&$ C_{d_W}(10\, \text{TeV})$ &$ C_{Wq}(10\, \text{TeV})$&$ C_{Zq}(10\, \text{TeV})$  \\
 \hline
 $C_{ u,\,d}\left(M_W\right)$ &$ \lbrace3.2,\,1.3\rbrace$ & $\lbrace6.4,\, 13 \rbrace$&$ \lbrace 11,\,19\rbrace $& $\lbrace 0.65,\,1.3\rbrace$ &$\lbrace -25,\, -29\rbrace$&$\lbrace -4.9,\, 8.8\rbrace$\\
$ \tilde{C}_{ u,\,d}\left(M_W\right) $& $-$ & $-$ &$-$&$-$ &$\lbrace-3.6, -3.6 \rbrace$&$\lbrace -0.81 ,\, 1.5\rbrace$ \\ 
 \hline
\end{tabular}
\end{center}
\label{TableMTMWHeavy} 
\end{table}

Employing all the anomalous dimensions, the finite contribution from $O_{d_W}$, and using
 \cite{Beringer:1900zz}
\bea \al_w(M_W) \simeq \frac{1}{128}, \qquad s_w\sq(M_W)  \simeq 0.23,\eea
we calculate the induced quark (C)EDMs at $M_W$. The results for $\MT = 1,\, 10 \,\text{TeV}$ are given in Table \ref{TableMTMWHeavy}. The quark EDM entries for $C_B$, $C_W$, and $C_{WB}$ for $\MT =10\,\mathrm{TeV}$ are almost twice as large as those for $\MT=1\,\mathrm{TeV}$. This can be understood from the logarithmic dependence on $\MT$, while, simultaneously, QCD corrections suppress the contributions. The same holds for the quark CEDM entries for $C_{W_q}$ and $C_{Z_q}$. The quark EDM entries for these last two operators are more complicated because they get additional contributions when the induced quark CEDM is evolved to lower energies.

The quark (C)EDMs can be run down to $\MQCD$ using results from Section \ref{unsup}. Following Section \ref{bounds} we use the neutron EDM limit to set bounds on the couplings of the heavy-boson operators. Using the same estimates as before, we obtain the results in Table \ref{TableHeavyBounds}.

\begin{table}[t!]
\caption{\small Bounds on the couplings of various $\slashPT$ operators involving heavy bosons. All entries are dimensionless.}
\begin{center}\footnotesize
\begin{tabular}{||c|cc||}
\hline 
& $\MT = 1\, \text{TeV}$& $\MT = 10\, \text{TeV}$\\
\hline 
 $(\MT\sq)C_B\left(\MT\right)$ &$\leq 8.1\cdot 10^{-2}$&$\leq 4.6$ \\
 $(\MT\sq)C_W\left(\MT\right) $& $\leq 1.9\cdot 10^{-2}$&$\leq 1.1$\\
 $(\MT\sq)C_{WB}\left(\MT\right) $&$\leq 1.3\cdot 10^{-2}$&$\leq 0.74$\\
 $(\MT\sq) C_{d_W}\left(\MT\right)$ &$\leq 0.11$&$\leq 11$\\
 $(\MT\sq) C_{Wu,\,d}\left(\MT\right)$&$\leq\lbrace{1.0,\, 0.84\rbrace}\cdot 10^{-2}$&$\leq\lbrace{0.53,\,0.45\rbrace}$\\
 $(\MT\sq) C_{Zu,\,d}\left(\MT\right)$ &$\leq\lbrace{5.3,\, 2.8\rbrace}\cdot 10^{-2}$&$\leq\lbrace{2.7,\, 1.4\rbrace}$\\
 \hline
\end{tabular}
\end{center}
\label{TableHeavyBounds} 
\end{table}

The limits on the different operators are all of similar size apart from those on $C_{d_W}$ which are somewhat weaker. A comparison with Table \ref{TableMTMQCDBounds} shows that the bounds on the heavy-boson operators are approximately an order of magnitude weaker than the bounds on the quark (C)EDMs. While they are a factor of $\Or(10^{2}$-$10^{3})$ weaker than the limits on $\mathrm{ Im} \,Y^{\prime q}$, $\mathrm{ Im}\, \Xi_{1}$, and $\mathrm{ Im} \,\Sigma_{1,8}$.
We mention again that these bounds cannot simply be used to constrain $\MT$. The effective couplings will depend on dimensionless quantities coming from the high-energy theory. In fact, in theories for which the results of Ref.\ \cite{Arzt:1994gp} hold all operators discussed in this section cannot be generated at tree level and so their couplings will be suppressed by loop factors $\sim 1/(4\pi)^2 \simeq 10^{-2}$. Since the bounds at $1\,\mathrm{TeV}$, let alone $10\,\mathrm{TeV}$, are all of that level or weaker, for such theories the neutron EDM does not significantly constrain the coupling constants of the dimension-six operators at this scale.

Some of the operators studied here have been under recent investigation as they can modify the $h\rightarrow \g\g$ rate \cite{McKeen:2012av, Grojean:2013kd, Fan:2013qn}. In this context, Ref. \cite{Fan:2013qn} used the electron EDM limit to set bounds on $O_W$, $O_B$, $O_{WB}$, and $O_{d_W}$. (Note that the obtained limits are, for all four operators, stronger than limits obtained from accelerator-based experiments \cite{Abdallah:2008sf}). The obtained constraints are approximately an order of magnitude stronger than those derived here. The main difference is due to the fact that the electron EDM limit  \cite{Hudson:2011zz} is about $30$ times stronger than the neutron EDM limit, although this effect is softened by the smallness of the electron mass with respect to the light-quark mass. Furthermore, the electron EDM is not subject to QCD corrections which suppress the quark EDMs by a small amount when they are evolved to lower energies. Overall, the electron and neutron EDM searches are complementary because they are sensitive to different $\slashPT$ sources. For example, if the dominant $\slashPT$ source is the operator $O_W$ the electron and neutron EDM would be of approximately the same size, whereas if, say, the gluon CEDM is the dominant source, the neutron EDM is expected to be much larger. This illustrates that measurements on different systems are needed to disentangle the fundamental $\slashPT$ mechanism.

\section{Discussion and conclusion}\label{conclusion}
A measurement of a nonzero EDM in any of the upcoming experiments would be a major breakthrough. 
Within the SM, a non-zero hadronic or nuclear EDM at the current experimental accuracies, can only originate in the $\tb$ term because $P$ and $T$ violation from the quark-mixing matrix is simply too small. However, the severe suppression of $\tb$ leaves room for physics beyond the SM. The effects of new physics can be parametrized by effective higher-dimensional operators which start at dimension six \cite{Buchmuller:1985jz}. In this article we have performed a systematic study of the dimension-six operators relevant to flavor-diagonal $P$ and $T$ violation in hadronic and nuclear systems. In particular, we have investigated the evolution of these operators from the energy scale where they originate ($\MT$), assumed to be significantly larger than the electroweak scale ($\MW$), to the QCD scale ($\MQCD \sim\,1\,\mathrm{GeV}$). 

The effective operators appearing at $\MT$ contain SM fields only and obey the SM $SU_c(3)\times SU_L(2)\times U_Y(1)$ gauge symmetries. The couplings of these operators are all proportional to two inverse powers of $\MT$. Gauge invariance puts strong constraints on the form of the effective operators. For example, only two $\slashPT$ four-quark operators containing first-generation quarks, \textit{i.e.} the FQPS operators, are allowed. Operators such as the quark EDM and CEDM which have canonical dimension five at low energies are actually dimension-six operators in disguise. Their chiral-symmetry-breaking properties forces them to be coupled to the Higgs field at high energies in order to preserve the SM gauge symmetries. Although at lower energies the Higgs field takes on its vev, the operators still scale as $\MT^{-2}$  \cite{DeRujula:1990db}.

In order to study the effects of the dimension-six operators on low-energy observables, such as EDMs, it is necessary to evolve the operators to these lower-energy scales. In the process heavy SM fields decouple and must be integrated out. Simultaneously, it is necessary to include the effects of QCD and, in some cases, electroweak renormalization-group running. While doing so, we have found it convenient to divide the dimension-six operators at $\MT$ into two sets. The division is based on the size of the induced low-energy $\slashPT$ operators. The first set contains the dimension-six operators which induce low-energy operators that also scale as $\MT^{-2}$. That is, the operators obtain no additional suppression when evolved to lower energies apart from minor suppressions due to QCD corrections. The second set contains the remaining operators which are suppressed by SM factors when run to $\MQCD$. Of course, EDM limits put the strongest constraints on operators in the first set.

\paragraph{Set 1:}

Operators in the first set have been discussed in Sec.\ \ref{unsup}. The set consists of operators which contain only light fields such as the gluon CEDM or the FQPS operators. The shape of these operators stays the same all the way from $\MT$ to $\MQCD$ and, apart from QCD corrections, their scalings are not altered. Other operators in the first set are the quark EDMs and CEDMs. At high energies these operators couple to a Higgs field, but these can be replaced by their vev and there are no suppression factors  at low energies. The next operator describes a derivative interaction among right-handed quarks and two Higgs fields (see Eq.\ \eqref{qqWlr}). When the Higgs field takes on its vev, an interaction among right-handed quarks and a $W$ boson remains.
Integrating out the $W$ boson induces an additional $\slashPT$ four-quark operator  (the FQLR operator) $\cite{Ng:2011ui}$. The $W$ exchange suppresses the operator by $M_W^{-2}$, but this is compensated by the two powers of $v$ appearing in the operator. Finally, the first set contains $\slashPT$ quark-Higgs and gluon-Higgs interactions. When the Higgs field takes its vev, the resulting terms renormalize, respectively, the SM Yukawa couplings and the $\tb$ term. Via loop diagrams the quark-Higgs \cite{Barr:1990vd} and gluon-Higgs interactions induces contributions to the quark (C)EDM. The associated loop suppressions are not very stringent and we therefore keep these operators in the first set. 

The low-energy form of the operators in the first set has been derived in Sec.\ \ref{sec32}. Around $\MQCD$ there are seven relevant $\slashPT$ operators among light quarks, gluons, and photons. They consist of the quark EDM, the quark and gluon CEDM, and four four-quark operators. The first three of these operators have been extensively considered in the literature. The four-quark operators are less often taken into account, possibly due to the belief that they should be proportional to the quark masses and are therefore effectively higher-dimensional operators in disguise. Although this arguments holds for some four-quark operators it does not hold for the FQPS and FQLR operators which genuinely scale as $\MT^{-2}$ and are in general not suppressed compared to the dipole operators. Model-independent studies of hadronic and nuclear EDMs due to beyond-the-SM physics should use the seven operators at $\MQCD$ as a starting point. Only model-dependent statements can single out particular operators in this set.

The FQPS and FQLR operators have very different origins. As mentioned, the FQPS operators conserve the SM gauge symmetries and appear directly at $\MT$. There is, \textit{a priori}, no link between the coupling constants of the color-singlet and color-octet FQPS operators since both are independently allowed by the gauge symmetries. Of course, such a link can exist in specific models of new physics. QCD corrections particularly affect the color-singlet operator when it is evolved to $\MQCD$ as can be seen from Eqs.\ \eqref{FQPSlow1} and \eqref{FQPSlow10}.
The FQLR operators do not conserve the $SU_2(L)$ gauge symmetry and are therefore not allowed at $\MT$. Nevertheless, the color-singlet FQLR is generated below the electroweak scale after a $W$-boson exchange \cite{Ng:2011ui}. The color-octet FQLR operator is only induced when its color-singlet partner is evolved to lower energies. The two operators are therefore not independent and both depend on the same coupling constant $\Xi_1$ appearing in Eq.\ \eqref{qqWlr} \cite{Lagdim6}. Eq.\ \eqref{FQLRlow} shows that at $\MQCD$ the coefficient of the color-octet FQLR operator is even slightly larger than that of the color-singlet operator. 

In Sec.\ \ref{bounds} we have used the neutron EDM limit to set bounds on the seven operators at $\MQCD$ using $\chi$PT calculations \cite{Vri11a, Lagdim6}. By use of the results in Secs. \ref{sec31} and \ref{sec32} these limits are translated to limits on all dimension-six operators in the first set at energies around $\MT$. This makes it possible to quickly obtain restrictions imposed by the neutron EDM on specific high-energy models. After performing a matching calculation between the high-energy model and the effective dimension-six operators, these restrictions can be read off immediately. 

There are interesting proposals to measure the EDMs of light nuclei in storage rings with high accuracy \cite{storageringexpts}. 
Such an experimental program could give important complementary information on the dominant $\slashPT$ mechanism. In Refs. \cite{Vri11a, Vri12, Bsaisou:2012rg, Lagdim6} a strategy has been proposed to (partially) disentangle the $\slashPT$ sources at the energy scale $\MQCD$. For example, a deuteron EDM significantly larger than the sum of the neutron and the proton EDM would point towards a quark CEDM or the FQLR operators \cite{Leb04, Vri12}. On the other hand, if the quark EDM is dominant one would expect the deuteron EDM to be close to the sum of the nucleon EDMs \cite{Vri11a}. This strategy combined with the results obtained here can be used to (partially) disentangle the various $\slashPT$ sources at high energies, if any exist.

\paragraph{Set 2:} The second set of dimension-six operators at $\MT$ consists of the remaining operators which do suffer from additional suppression when evolved to lower energies. We first discuss $\slashPT$ quark-Higgs interactions which already appeared in the first set because they induced relatively large corrections to the quark CEDM via loops. At tree-level the quark-Higgs couplings induce $\slashPT$ four-quark operators which are different from the FQPS and FQLR operators. However, the couplings of these operators scale as $(m_q/v)\MT^{-2}$ and are thus additionally suppressed by the ratio of the light-quark mass to the Higgs vev. It is interesting that when the quark-Higgs operators are evolved to energies around $\MQCD$, all ten possible $\slashPT$ four-quark operators among up and down quarks are induced, albeit with only two independent couplings and the coefficients of three operators are an order of magnitude smaller than the others. The QCD evolution of these operators was studied in a different basis in Refs. \cite{ An:2009zh, Hisano:2012cc}. The set contains the FPQS and FQLR operators but also six new structures. However, because the quark-Higgs operators induce a larger contribution to the quark CEDM, the additional six four-quark operators can in general be neglected. 

The other dimension-six operators in the second set describe $\slashPT$ interactions involving electroweak gauge and/or Higgs bosons. The operators consist of the quark weak dipole moments and $\slashPT$ interactions among three gauge bosons and/or two gauge bosons and a Higgs boson. The quark weak dipole moments are similar to the quark EDM but the photon is replaced by a $W$ or $Z$ boson. In fact, as seen from Eqs.\  \eqref{dq}, \eqref{qz}, and \eqref{qw} these operators are related by gauge symmetry. The purely bosonic interactions originate in four independent gauge-invariant operators \cite{Buchmuller:1985jz}. 

All operators contain at least one heavy field which decouples below the electroweak scale generating contributions to the quark EDMs and CEDMs via electroweak one-loop diagrams \cite{DeRujula:1990db}. These contributions are typically suppressed by $\alpha_w/4\pi$. At the one-loop level there are no dimension-six four-quark operators generated. In Sec.\ \ref{heavyboson} we have calculated the induced quark (C)EDMs and used the neutron EDM limit to set bounds on the various dimension-six operators.  These bounds are weaker than the bounds on the operators in the first set, but by how much depends on the particular operator. 

The purely bosonic dimension-six operators induce, apart from quark EDMs, also contributions to the electron EDM \cite{Fan:2013qn} via very similar diagrams. The limits obtained from the electron EDM are stronger than from the neutron EDM (both are stronger than limits from accelerator experiments \cite{Fan:2013qn, Abdallah:2008sf}) but this might change depending on which EDM limit is improved first.  What is interesting is that the bosonic  operators induce an electron and neutron EDM of similar size, while the other dimension-six operators discussed in this paper induce a significantly larger neutron EDM. Observation of such a pattern in future experiments could provide a hint towards the fundamental source of $P$ and $T$ violation.
\\
\\
Finally, we should say that our set of dimension-six operators around $\MT$ is far from complete since we did not include operators involving heavier quarks. Allowing such fields would increase the number of operators significantly, especially if one allows for generation-changing operators. Most importantly, we did not consider operators involving strange quarks which are still present in the EFT at $\MQCD$ and can have important consequences for hadronic and nuclear EDMs. We leave a study of operators involving strangeness to future work. In general, we expect operators involving heavier quarks than up, down, and strange to give suppressed contributions to operators around $\MQCD$ due to the need of integrating out the heavier quarks. However, exceptions to this argument exist, see, for example, discussions in Refs. \cite{Pospelov:2005pr, Hisano:2012cc}. 

In conclusion, we have performed a systematic study of parity- and time-reversal violating operators of dimension six which originate in physics beyond the SM.  Beginning at the high-energy scale where these operators originate and their forms are constrained by gauge invariance, we have evolved the operators down to the electroweak scale and subsequently to hadronic scales. We have derived a set of operators which is expected to dominate hadronic and nuclear EDMs due to physics beyond the SM. Furthermore, we have obtained quantitative relations between these operators and the original dimension-six operators at the high-energy scale.

\section*{Acknowledgements}
We thank D. Boer, E. Mereghetti, R. Timmermans, and U. van Kolck for many discussions and comments on the manuscript. This research was supported by the Dutch Stichting FOM
under programs 104 and 114 and in part by the DFG and the NSFC through funds provided to
the Sino-German CRC 110 ``Symmetries and the Emergence of Structure in QCD''  (JdV).

\end{document}